\documentclass[aps,prd,showpacs,floatfix,eqsecnum,amsmath,amssymb,nofootinbib,twocolumn]{revtex4}

\usepackage{graphicx}

\newcommand{\be}{\begin{equation}}
\newcommand{\ee}{\end{equation}}
\newcommand{\bse}{\begin{subequations}}
\newcommand{\ese}{\end{subequations}}

\newcommand{\barA}{{\bar{A}}}
\newcommand{\btau}{\boldsymbol{\tau}}
\newcommand{\btheta}{\boldsymbol{\theta}}
\newcommand{\bxi}{\boldsymbol{\xi}}
\newcommand{\bzeta}{\boldsymbol{\zeta}}

\newcommand{\dst}{\displaystyle}

\newcommand{\F}{{\mathcal F}}
\newcommand{\Fo}{{\mathcal F}_0}

\newcommand{\md}{\mathrm{d}}
\newcommand{\mi}{\mathrm{i}}

\newcommand{\pa}{\partial}
\newcommand{\phim}{\phi_\mathrm{m}}
\newcommand{\phis}{\phi_\mathrm{s}}

\newcommand{\sfD}{{\mathsf{D}}}
\newcommand{\sfM}{{\mathsf{M}}}
\newcommand{\sfO}{{\mathsf{O}}}
\newcommand{\sfS}{{\mathsf{S}}}
\newcommand{\sfU}{{\mathsf{U}}}

\newcommand{\To}{{T_\mathrm{o}}}
\newcommand{\tav}[1]{\left\langle#1\right\rangle}
\newcommand{\tb}{t_\mathrm{b}}

\newcommand{\ve}{\varepsilon}

\begin{document}

\title{Data analysis of gravitational-wave signals
from spinning neutron stars.\\ V.\ A narrow-band all-sky search}

\author{Pia Astone}
\affiliation{Istituto Nazionale di Fisica Nucleare,
(INFN)-Rome I, 00185 Rome, Italy}

\author{Kazimierz M.\ Borkowski}
\affiliation{Centre for Astronomy,
Nicolaus Copernicus University,
Gagarina 11, 87-100 Toru\'n, Poland}

\author{Piotr Jaranowski}
\affiliation{Faculty of Physics,
University of Bia{\l}ystok,
Lipowa 41, 15-424 Bia{\l}ystok, Poland}

\author{Andrzej Kr\'olak}
\affiliation{Institute of Mathematics,
Polish Academy of Sciences,
\'Sniadeckich 8, 00-950 Warsaw, Poland}

\author{Maciej Pietka}
\affiliation{Faculty of Physics,
University of Bia{\l}ystok,
Lipowa 41, 15-424 Bia{\l}ystok, Poland}

\begin{abstract}

We present theory and algorithms to perform an all-sky coherent search
for periodic signals of gravitational waves in narrow-band data of a detector.
Our search is based on a statistic, commonly called the $\F$-statistic,
derived from the maximum-likelihood principle in Paper I of this series.
We briefly review the response of a ground-based detector
to the gravitational-wave signal from a rotating neuron star
and the derivation of the $\F$-statistic.
We present several algorithms to calculate efficiently this statistic.
In particular our algorithms are such that one can take advantage
of the speed of fast Fourier transform (FFT) in calculation of the $\F$-statistic.
We construct a grid in the parameter space such that the nodes of the grid
coincide with the Fourier frequencies.
We present interpolation methods that approximately convert the two integrals
in the $\F$-statistic into Fourier transforms
so that the FFT algorithm can be applied in their evaluation.
We have implemented our methods and algorithms into computer codes
and we present results of the Monte Carlo simulations performed to test these codes.

\end{abstract}

\pacs{95.55.Ym, 04.80.Nn, 95.75.Pq, 97.60.Gb}

\maketitle

\section{Introduction}
\label{Sec:Intro}

Periodic gravitational-wave signals like those originating from rotating neutron
stars
are an important class of sources that can be detected by currently operating
ground-based detectors.
Several methods were developed to search for such sources and several searches
were performed.
This paper continues the series of papers \cite{JKS98,JK99,JK00,ABJK02}
devoted to studies of data analysis tools and algorithms
needed to perform an all-sky coherent search for quasiperiodic gravitational
waves.

The search presented in the current paper is based on the maximum-likelihood
statistic called the $\F$-statistic
that we have derived in the Paper I \cite{JKS98} of this series.
It is known that the coherent search for long observation time needed to detect
weak gravitational-wave signals
from rotating neutron stars are computationally prohibitive (see \cite{BCCS98}
and Paper III of this series \cite{JK00}).
Promising strategies are hierarchical semi-coherent methods.
In these methods data is broken into short segments. In the first stage each
segment is analyzed using
the $\F$-statistic and in the second stage the $\F$-statistics from the short
segments are combined using a certain algorithm.
There are several methods proposed for the second stage:
search for coincidences among candidates from short duration segments
\cite{LIGO07Fstat,LIGO09EHa},
stack-slide method \cite{BC99}, power flux method
\cite{LIGO08powerflux,LIGO09powerflux},
Hough transform method \cite{PAFS97,PS99,KSPSFP04,LIGO05Hough,LIGO08powerflux}.
Recently an optimal method for the second stage has been found, the {\em global
correlation coordinate} method \cite{P08,AP09},
which exploits global parameter space correlations in the coherent detection statistic.
In our paper we shall present methods to optimize the first, coherent stage of a hierarchical method.

The techniques presented in this paper were used in the analysis of NAUTILUS bar
detector data \cite{Astone2008}
and are presently used in the analysis of the VIRGO data.
Alternative techniques for the coherent stage based on the $\F$-statistic
and their application to the real data can be found in Refs.\
\cite{Astone2005,LIGO07Fstat,LIGO09EHa,LIGO09EHb}

The paper is organized as follows.
In Sec.\ \ref{Sec:Pio} we present the noise-free response of a ground-based
detector
to a gravitational-wave signal from a rotating neutron star.
This response was derived and discussed in detail in Papers I \cite{JKS98} and
IV \cite{ABJK02} of our series.
In Sec.\ \ref{Sec:Kro1} we present data analysis tools to perform coherent
search of the data
for a gravitational-wave signal given in Sec.\ \ref{Sec:Pio}.
In Sec.\ \ref{sSec:Kro1a} we present the $\F$-statistic that was derived in
Paper I.
We limit ourselves to the case when the observation time is an integer multiple
of one sidereal day.
This simplifies some general formulas considerably.
In Sec.\ \ref{sSec:Kro1b} we introduce a simplified approximate model for a
periodic gravitational-wave signal.
This approximate signal has the constant amplitude
and its phase is parameterized in such a way that it is a linear function of the
parameters.
For such a signal the Fisher matrix is constant and consequently
it is independent of the values of the signal's parameters.
In Sec.\ \ref{sSec:false} we briefly review calculation of the false alarm probability.
Section \ref{Sec:Mac} is devoted to construction of the grid of templates in the
parameter space.
The grid solves a certain covering problem with a constraint.
Our constraint is that the nodes of the grid coincide with the Fourier
frequencies.
This allows to use the fast Fourier transform (FFT) algorithm to compute the
$\F$-statistic at grid nodes,
what greatly accelerates the calculation.
In Sec.\ \ref{Sec:Kaz} we describe our package \texttt{Top2Bary}
that is used to calculate the position and the velocity of the detector located
on the Earth
with respect to the solar system barycenter.
In Sec.\ \ref{sSec:Kaza} we introduce various concepts and definitions used in
the astrometry
and in Sec.\ \ref{sSec:Kazb} we describe the content of our package which is a
set of \textsc{fortran} routines.
In Sec.\ \ref{Sec:Kro} we present various approximations
that we use in the calculation of the $\F$-statistic in order to speed up
computations.
In Sec.\ VI~A we discuss resampling of the time series to the barycenter
that we need to perform before we can apply the FFT.
We develop two algorithms: one slow and very accurate and the other fast but less accurate.
We compare the two algorithms using the signal from Sec.\ \ref{Sec:Pio}.
In Sec.\ VI~B we describe interpolation of the FFT in the Fourier domain.
This interpolation method allows to obtain efficiently
an FFT that is twice as fine as the FFT of original data.
In Sec.\ VI~C we describe the Nelder-Mead algorithm that we use to find
accurately the maximum of the $\F$-statistic.
In Sec.\ \ref{Sec:Kro3} we perform a number of Monte Carlo simulations of the computer code
where we have implemented the methods and algorithms from Secs.\ III--VI.
In our simulations we investigate how well we estimate the parameters of the signal
in comparison to the Cram\'er-Rao bound.

\section{Response of a detector to a~periodic gravitational wave}
\label{Sec:Pio}

The dimensionless noise-free response $h$ of a gravitational-wave detector
to a weak plane gravitational wave in the {\em long wavelength approximation}
[i.e.,\ when the size of the detector is much smaller
than the reduced wavelength $\lambda/(2\pi)$ of the wave]
can be written as the linear combination
of the two independent wave polarizations $h_+$ and $h_\times$,
\be
\label{resp}
h(t) = F_+(t) h_+(t) + F_\times(t) h_\times(t),
\ee
where $F_+$ and $ F_\times$ are the detector's beam-pattern functions,
which are of the form
\bse
\label{patt}
\begin{align}
F_+(t) &= \sin\zeta\big(a(t)\cos 2\psi+b(t)\sin 2\psi\big),
\\[1ex]
F_\times(t) &= \sin\zeta\big(b(t)\cos 2\psi-a(t)\sin 2\psi\big).
\end{align}
\ese
The beam-patterns $F_+$ and $ F_\times$ are linear combinations of $\sin2\psi$
and $\cos2\psi$,
where $\psi$ is the polarization angle of the wave.
For interferometric detectors the angle $\zeta$
is the angle between the interferometer arms (usually $\zeta=90^\circ$)
whereas for the case of bars one has to put $\zeta=90^\circ$.
The functions $a(t)$ and $b(t)$ are amplitude modulation functions,
which depend on the location of the detector on the Earth
and on the position of the gravitational-wave source in the sky
(described in the celestial coordinate system
by the right ascension $\alpha$ and the declination $\delta$ of the source).
They are periodic functions of time with the period of one sidereal day.
Analytic form of the functions $a(t)$ and $b(t)$ depends on the type of the
detector;
for the case of bar detectors they are explicitly given in Eqs.\ (A11) of Ref.\
\cite{ABJK02},
whereas for interferometric detectors they can be found in Eqs.\ (12) and (13)
of Ref.\ \cite{JKS98}.

We are interested in periodic waves,
for which the wave polarization functions are of the form
\bse
\label{hphc}
\begin{align}
h_+(t) &= h_{0+} \cos(\phi(t)+\phi_0),
\\[1ex]
h_\times(t) &= h_{0\times} \sin(\phi(t)+\phi_0),
\end{align}
\ese
where $h_{0+}$ and $h_{0\times}$ are constant amplitudes of the two
polarizations
and $\phi(t)+\phi_0$ is the phase of the wave
(with $\phi_0$ being the initial phase of the waveform).
The amplitudes $h_{0+}$ and $h_{0\times}$ depend
on the physical mechanism generating gravitational radiation.
E.g., if a neutron star is a triaxial ellipsoid rotating around
a principal axis with frequency $f$, then these amplitudes are
\bse
\label{h0iota}
\begin{align}
h_{0+} &= \frac{1}{2} h_0 (1 + \cos^2\iota),
\\[1ex]
h_{0\times} &= h_0 \cos\iota,
\end{align}
\ese
where $\iota$ is the angle between the star's angular momentum vector
and the direction from the star to the Earth, and the amplitude $h_0$ is given
by
\be
\label{h0}
h_0 = \frac{16\pi^2G}{c^4} \frac{\epsilon I f^2}{r}.
\ee
Here $I$ is the star's moment of inertia with respect to the rotation axis,
$r$ is the distance to the star, and $\epsilon$ is the star's ellipticity
defined by $\epsilon=|I_1-I_2|/I$, where $I_1$ and $I_2$ are moments of inertia
with respect to the principal axes orthogonal to the rotation axis.

We further assume that the gravitational waveform
given by Eqs.\ \eqref{resp}--\eqref{hphc}
is almost monochromatic around some angular frequency $\omega_0$,
which we define as instantaneous angular frequency
evaluated at the solar system barycenter (SSB) at $t=0$.
The phase modulation function $\phi(t)$ for such waveform
is approximately given by
\be
\label{pha3}
\phi(t) = \sum_{k=0}^{s}\omega_k \frac{t^{k+1}}{(k+1)!}
+ \frac{{\bf n}_0\cdot{\bf r}_{\rm d}(t)}{c} \sum_{k=0}^{s}
\omega_k\frac{t^k}{k!},
\ee
where $\omega_k$ ($k=1,2,\ldots,s$) is the $k$th time derivative
of the instantaneous angular frequency at the SSB evaluated at $t=0$,
$\mathbf{n}_0$ is the constant unit vector
in the direction of the star in the SSB reference frame
(it depends on the right ascension $\alpha$ and the declination $\delta$ of the
source),
and $\mathbf{r}_\mathrm{d}$ is the vector joining the SSB with the detector.
Approximations that lead to Eq.\ \eqref{pha3} are discussed in detail
in Sec.\ II~B and Appendix~A of Paper~I.

Let us associate the following coordinate system with the SSB reference frame.
The $x$ axis of the system is parallel to the $x$ axis of the celestial
coordinate system,\footnote{In the celestial coordinate system
the $z$ axis coincides with the Earth's rotation axis
and points toward the North pole, the $x$ and $y$ axes lie in the Earth's
equatorial plane
with the $x$ axis pointing toward the vernal point.}
the $z$ axis is perpendicular to the ecliptic
and has direction of the orbital angular momentum vector of the Earth.
In this SSB coordinate system the vector $\mathbf{n}_0$ has the components
\be
\label{pha4}
\mathbf{n}_0=\left(\begin{array}{ccc}
1 &    0     & 0       \\
0 &  \cos\ve & \sin\ve \\
0 & -\sin\ve & \cos\ve
\end{array}\right)
\left(\begin{array}{c}
\cos\alpha\cos\delta\\ \sin\alpha\cos\delta\\ \sin\delta
\end{array}\right),
\ee
where $\ve$ is the obliquity of the ecliptic.
The position vector ${\bf r}_{\rm d}$ of the detector
with respect to the SSB has the components
\be
\label{pha5}
{\bf r}_{\rm d}
= \left(\begin{array}{c}
R^x_\mathrm{ES} \\[1ex] R^y_\mathrm{ES} \\[1ex] 0
\end{array}\right)
+ \left(\begin{array}{ccc}
1 &    0     & 0       \\
0 &  \cos\ve & \sin\ve \\
0 & -\sin\ve & \cos\ve
\end{array}\right)
\left(\begin{array}{c}
R^x_\mathrm{E} \\[1ex] R^y_\mathrm{E} \\[1ex] R^z_\mathrm{E}
\end{array}\right),
\ee
where $(R^x_\mathrm{ES},R^y_\mathrm{ES},0)$
are the components of the vector joining the SSB
with the center of the Earth in the SSB coordinate system,
and $(R^x_\mathrm{E},R^y_\mathrm{E},R^z_\mathrm{E})$ are the
components of the vector joining the center of the Earth
and the detector's location in the celestial coordinate system.
Making use of Eqs.\ \eqref{pha4} and \eqref{pha5} one can obtain
the explicit formula for the scalar product
${\bf n}_0\cdot{\bf r}_{\rm d}(t)$:
\begin{align}
{\bf n}_0 \cdot {\bf r}_{\rm d}(t)
&= \cos\alpha\cos\delta\big(R^x_\mathrm{ES}(t) + R^x_\mathrm{E}(t)\big)
\nonumber\\[1ex]&\quad
+ (\sin\alpha\cos\delta\cos\ve + \sin\delta\sin\ve) R^y_\mathrm{ES}(t)
\nonumber\\[1ex]&\quad
+ \sin\alpha\cos\delta\,R^y_\mathrm{E}(t) + \sin\delta\,R^z_\mathrm{E}(t).
\end{align}

The phase $\phi$ of the waveform given by Eq.\ \eqref{pha3}
depends on the angular frequency $\omega_0$,
the $s$ spin-down parameters $\omega_k$ $(k=1,\ldots,s)$,
and on the angles $\alpha$, $\delta$ (through the vector $\mathbf{n}_0$).
We call the parameters $(\omega_0,\omega_1,\ldots,\omega_s,\alpha,\delta)$
the {\em intrinsic} parameters and the remaining ones
$(h_{0+},h_{0\times},\phi_0,\psi$) the {\em extrinsic} (or {\em amplitude})
parameters.
As we shall see in the following section
we only need to search for signals over the intrinsic parameter space.
The whole signal $h$ depends on $s+7$ unknown parameters:
$(h_{0+},h_{0\times},\phi_0,\psi,\alpha,\delta,\omega_0,\omega_1,\ldots,
\omega_s)$.

The response function $h$ depends
on the position of the detector with respect to the SSB.
This position can be determined with a great accuracy
using JPL Planetary and Lunar Ephemerides DE405/LE405
as described in Sec.\ \ref{Sec:Kaz}.
The dominant term in the phase $\phi(t)$ is $\omega_0\,t$;
typical gravitational-wave frequency $f_0:=2\pi/\omega_0$
is contained in the range from a few Hz to a few kHz.
The gravitational-wave signal from a rotating neutron star
is a nearly periodic signal that is weakly amplitude and phase modulated
due to the intrinsic variation of star's rotation frequency
and the motion of the detector with respect to the star.
Moreover the amplitude of this signal is expected to be very small.
Consequently detection of the signal requires observation time $T_\mathrm{o}$
that is very long with respect to the gravitational-wave period
$P_0:=2\pi/\omega_0$.

Combining Eqs.\ \eqref{resp}--\eqref{hphc} together one can decompose
the response $h$ into linear combination of four time-dependent components:
\be
\label{sig}
h(t) = \sum^4_{i=1} A_{i}\,h_{i}(t),
\ee
where the functions $h_i$ ($i=1,\dots,4$) are of the form
\begin{align}
\label{eq:ab}
\begin{array}{c}
h_1(t) = a(t)\cos\phi(t), \hspace{3mm} h_2(t) = b(t)\cos\phi(t),
\\[2ex]
h_3(t) = a(t)\sin\phi(t), \hspace{3mm} h_4(t) = b(t)\sin\phi(t),
\end{array}
\end{align}
and the four constant amplitudes $A_{i}$ ($i=1,\dots,4$) are given by
\be
A_i = \barA_i \sin\zeta,\quad i=1,\dots,4,
\ee
where
\bse
\label{eq:ampone}
\begin{align}
\barA_1 &= h_{0+}\cos2\psi\cos\phi_0 - h_{0\times}\sin2\psi\sin\phi_0,
\\[1ex]
\barA_2 &= h_{0+}\sin2\psi\cos\phi_0 + h_{0\times}\cos2\psi\sin\phi_0,
\\[1ex]
\barA_3 &= - h_{0+}\cos2\psi\sin\phi_0 - h_{0\times}\sin2\psi\cos\phi_0,
\\[1ex]
\barA_4 &= - h_{0+}\sin2\psi\sin\phi_0 + h_{0\times}\cos2\psi\cos\phi_0.
\end{align}
\ese

One can invert Eqs.\ \eqref{eq:ampone}
to obtain formulas for the parameters
$h_{0+}$, $h_{0\times}$, $\phi_0$, and $\psi$
as functions of the amplitudes $\barA_{i}$.
Let us introduce quantities
\bse
\begin{align}
A &:= \barA_1^2 + \barA_2^2 + \barA_3^2 + \barA_4^2,
\\[1ex]
D &:= \barA_1 \barA_4 - \barA_2 \barA_3.
\end{align}
\ese
Then the amplitudes $h_{0+}$ and $h_{0\times}$
can be uniquely determined from the relations
(we assume here, without loss of generality, that $h_{0+}>0$)
\bse
\begin{align}
h_{0+} &= \sqrt{\frac{1}{2}\Big(A + \sqrt{A^2 - 4D^2}\Big)},
\\[1ex]
h_{0\times} &= \mbox{sign}(D)
\sqrt{\frac{1}{2}\Big(A - \sqrt{A^2 - 4D^2}\Big)}.
\end{align}
\ese
The initial phase $\phi_0$ and the polarization angle $\psi$
can be obtained from the following equations:
\bse
\begin{align}
\tan2\phi_0 &= \frac{2(\barA_1 \barA_3 + \barA_2 \barA_4)}
{\barA_3^2 + \barA_4^2 - \barA_1^2 - \barA_2^2},
\\[1ex]
\tan4\psi &= \frac{2(\barA_1 \barA_2 + \barA_3 \barA_4)}
{\barA_1^2 + \barA_3^2 - \barA_2^2 - \barA_4^2}.
\end{align}
\ese
Also Eqs.\ \eqref{h0iota} can be solved
for the amplitude $h_0$ and the angle $\iota$. The result is
\bse
\begin{align}
h_0 &= h_{0+} + \sqrt{h_{0+}^2 - h_{0\times}^2},
\\[1ex]
\iota &= \arccos(h_{0\times}/h_0).
\end{align}
\ese

In the special case when the star's angular momentum vector
lies along the line of sight, $\cos\iota=\pm1$, and the number
of independent amplitude parameters is reduced to two.
In this situation Eqs.\ \eqref{h0iota} read
(upper sign is for $\cos\iota=+1$ and lower sign is for $\cos\iota=-1$)
\be
h_{0+} = h_0, \quad h_{0\times} = \pm h_0,
\ee
and Eqs.\ \eqref{eq:ampone} simplify then to
\bse
\label{ampdeg}
\begin{align}
\barA_1 &= h_0\cos(2\psi \pm \phi_0),
\\[1ex]
\barA_2 &= h_0\sin(2\psi \pm \phi_0),
\\[1ex]
\barA_3 &= \mp\barA_2,
\\[1ex]
\barA_4 &= \pm\barA_1.
\end{align}
\ese

\section{Maximum-likelihood filtering}
\label{Sec:Kro1}

\subsection{The $\F$-statistic}
\label{sSec:Kro1a}

The gravitational-wave signal $h$ given by Eqs.\ \eqref{sig} and \eqref{eq:ab}
will be buried in the noise of a detector.
We are thus faced with the problem of detecting the signal and estimating its
parameters.
A standard method is the method of {\em maximum-likelihood} (ML) detection
that consists of maximizing the likelihood function, which we shall denote by
$\Lambda$,
with respect to the parameters of the signal.
If the maximum of $\Lambda$ exceeds a certain threshold
calculated from the false alarm probability that we can afford,
we say that the signal is detected.
The values of the parameters that maximize $\Lambda$
are said to be the {\em maximum-likelihood estimators}
of the parameters of the signal.
The magnitude of the maximum of $\Lambda$
determines the probability of detection of the signal.

We assume that the noise $n$ in the detector is an additive,
stationary, Gaussian, and zero-mean continuous random process.
Then the data $x$ (if the signal $h$ is present) can be written as
\be
x(t) = n(t) + h(t).
\ee
The logarithm of the likelihood function has the form
\be
\label{loglr}
\ln\Lambda = (x|h) - \frac{1}{2}(h|h),
\ee
where the scalar product $(\,\cdot\,|\,\cdot\,)$ is defined by
\be
\label{SP}
(x|y) := \frac{2}{\pi}\, \Re \int^{\infty}_{0}
\frac{\tilde{x}(\omega)\tilde{y}^{*}(\omega)}{S_h(\omega)} \md\omega.
\ee
In Eq.\ \eqref{SP} tilde denotes the Fourier transform,
asterisk means complex conjugation,
$S_h$ is the {\em one-sided} spectral density of the detector's noise,
and $\Re$ denotes the real part of a complex expression.

We further assume that over the frequency bandwidth of the signal $h$
the spectral density $S_h$ is nearly constant and equal to $S_0=S_h(\omega_0)$,
where $\omega_0$ is the frequency of the signal measured at the SSB at $t=0$.
Then the scalar products entering Eq.\ \eqref{loglr} can be approximated by
\bse
\label{approx1}
\begin{align}
(x|h) &\approx \frac{2}{S_0}
\int^{T_\mathrm{o}}_{0} x(t)\,h(t)\,\md t,
\\[1ex]
(h|h) &\approx \frac{2}{S_0}
\int^{T_\mathrm{o}}_{0} \big(h(t)\big)^2\,\md t,
\end{align}
\ese
where $T_\mathrm{o}$ is the observation time,
and the observation interval is $\left\langle0,T_\mathrm{o}\right\rangle$.
It is useful to introduce the following notation
\be
\tav{x} := \frac{1}{T_\mathrm{o}} \int_0^{\To} x(t)\,\md t.
\ee
After applying this notation and making use of Eqs.\ (\ref{approx1}),
the log likelihood ratio from Eq.\ (\ref{loglr}) can be written as
\be
\label{loglr3}
\ln\Lambda \approx \frac{2\To}{S_0}
\left(\tav{xh}-\frac{1}{2}\tav{h^2}\right).
\ee

In Sec.\ III of Paper III we have analyzed in detail
the likelihood ratio for the general case of a signal
consisting of several narrow-band components.
Here we only summarize the results of Paper III
and adapt them to the case of our signal \eqref{sig}.
The signal $h$ depends linearly on four amplitudes $A_i$.
The likelihood equations for the ML estimators
$\widehat{A}_{i}$ of the amplitudes $A_i$ are given by
\be
\label{ampest}
\frac{\partial\ln\Lambda}{\partial A_{i}} = 0,
\quad i=1,\ldots,4.
\ee
One can easily find the explicit analytic solution to Eqs.\ \eqref{ampest}.
To simplify formulas we assume that {\em the observation time $\To$
is an integer multiple of one sidereal day}, i.e.,
$\To=n(2\pi/\Omega_\text{r})$ for some positive integer $n$,
where $\Omega_\text{r}$ is the rotational angular velocity of the Earth.
Then the time average of the product of the functions $a$ and $b$
[see Eqs.\ \eqref{patt}] vanishes,
$\tav{ab}=0$, and the analytic formulas for the ML estimators of the amplitudes
are given by
\be
\label{amle0}
\begin{array}{c}
\dst \widehat{A}_1 \approx 2 \frac{\tav{x h_1}}{\tav{a^2}}, \quad
\dst \widehat{A}_2 \approx 2 \frac{\tav{x h_2}}{\tav{b^2}},
\\[3ex]
\dst \widehat{A}_3 \approx 2 \frac{\tav{x h_3}}{\tav{a^2}}, \quad
\dst \widehat{A}_4 \approx 2 \frac{\tav{x h_4}}{\tav{b^2}}.
\end{array}
\ee
Explicit formulas for the time averages $\tav{a^2}$ and $\tav{b^2}$
can be found in Appendix B of Paper IV.

The reduced log likelihood function $\F$ or the $\F$-{\em statistic}
is the log likelihood function \eqref{loglr3}
with the amplitude parameters $A_i$
replaced by their estimators $\widehat{A}_i$.
By virtue of Eqs.\ \eqref{amle0} from Eq.\ \eqref{loglr3} one gets
\be
\label{OS}
\F \approx \frac{2}{S_0\To}
\left( \frac{|F_a|^2}{\tav{a^2}} + \frac{|F_b|^2}{\tav{b^2}} \right),
\ee
where
\bse
\label{Fab}
\begin{align}
\label{eq:Fa}
F_{a} &:= \int^{\To}_0 x(t)\, a(t) \exp[-\mi\phi(t)]\,\md t,
\\[1ex]
\label{eq:Fb}
F_{b} &:= \int^{\To}_0 x(t)\, b(t) \exp[-\mi\phi(t)]\,\md t.
\end{align}
\ese

The ML estimators of the signal's parameters are obtained in two steps.
Firstly, the estimators of the frequency, the spin-down parameters,
and the angles $\alpha$ and $\delta$ are obtained
by maximizing the functional $\F$ with respect to these parameters.
Secondly, the estimators of the amplitudes $A_{i}$
are calculated from the analytic formulas (\ref{amle0})
with the correlations $\tav{xh_i}$ evaluated
for the values of the parameters obtained in the first step.

\subsection{A linear model}
\label{sSec:Kro1b}

In this subsection we introduce a useful approximate model
of the gravitational-wave signal from a rotating neutron star.
The model relies on (i) neglecting all spin downs in the phase modulation
due to motion of the detector with respect to the SSB;
and (ii) discarding this component of the vector ${\bf r}_{\rm d}$
(connecting the SSB and the detector)
which is perpendicular to the ecliptic.
These approximations lead to the following phase of the signal:
\be
\label{philin}
\phi_\mathrm{lin}(t) = \sum_{k=0}^{s}\omega_k \frac{t^{k+1}}{(k+1)!}
+ \alpha_1 \mu_1(t) + \alpha_2 \mu_2(t),
\ee
where $\alpha_1$ and $\alpha_2$ are new constant parameters,
\bse
\label{eq:albe}
\begin{align}
\alpha_1 &:= \omega_0 (\sin\alpha\cos\delta\cos\ve + \sin\delta\sin\ve),
\\[1ex]
\alpha_2 &:= \omega_0 \cos\alpha\cos\delta,
\end{align}
\ese
and where $\mu_1(t)$ and $\mu_2(t)$ are known functions of time,
\bse
\begin{align}
\mu_1(t) &:= \frac{1}{c}\Big(R^y_\mathrm{ES}(t) + R^y_\mathrm{E}(t)\cos\ve\Big),
\\[1ex]
\mu_2(t) &:= \frac{1}{c}\Big(R^x_\mathrm{ES}(t) + R^x_\mathrm{E}(t)\Big).
\end{align}
\ese
We also neglect the slowly varying modulation of the signal's amplitude,
so finally we approximate the whole signal $h(t)$ by
\be
h(t) = A_0\cos\big(\phi_\mathrm{lin}(t)+ \phi_0\big),
\ee
where $A_0$ and $\phi_0$ are the constant amplitude and initial phase,
respectively.
The above signal model is called {\em linear}
because it has the property that its phase \eqref{philin}
is a linear function of the parameters.

It is convenient to represent the linear model
of the gravitational-wave signal in the following form
\be
\label{eq:siglin}
h(t;\btheta)
= A_0 \cos\bigg(\sum_{k=0}^M \xi_k m_k(t) + \phi_0\bigg),
\ee
where the vector $\btheta$ collects all the signal's parameters,
$\btheta:=(A_0,\phi_0,\bxi)$, with the vector $\bxi$ comprising
the parameters of the signal's phase,
$\bxi:=(\omega_0,\omega_1,\ldots,\omega_s,\alpha_1,\alpha_2)$,
so $\xi_k=\omega_k$ for $k=0,1,\ldots,s$, $\xi_{s+1}=\alpha_1$,
$\xi_{s+2}=\alpha_2$;
functions $m_k(t)$, $k=0,1,\ldots,s+2$,
are known functions of time $t$:
$m_k(t):=t^{k+1}/(k+1)!$ for $k=0,1,\ldots,s$,
$m_{s+1}(t):=\mu_1(t)$, and $m_{s+2}(t):=\mu_2(t)$;
finally, $M:=s+2$.

For the signal \eqref{eq:siglin} we will compute
the optimal signal-to-noise ratio $\rho$,
\be
\label{snr-def}
\rho := \sqrt{(h|h)},
\ee
and the components of the Fisher information matrix $\Gamma$,
\be
\label{Fm-def}
\Gamma_{k\ell}
:= \Big(\frac{\pa h}{\pa\theta_k}\Big|\frac{\pa h}{\pa\theta_\ell}\Big).
\ee
It is reasonable to assume that the observation time $\To$
is much longer than the period $P_0=2\pi/\omega_0$ of the gravitational wave
(typically $P_0\lesssim0.1$\,s and $\To\gtrsim1$\,day).
As a consequence
\be
\label{tav0}
\tav{\cos[n\phi_\mathrm{lin}(t)]} \approx 0,
\quad
\tav{\sin[n\phi_\mathrm{lin}(t)]} \approx 0,
\ee
for any positive integer $n$. Making use of these approximations
one easily computes from Eq.\ \eqref{snr-def} the signal-to-noise ratio,
\be
\label{snr}
\rho \approx A_0 \sqrt{\frac{\To}{S_0}},
\ee
and from Eq.\ \eqref{Fm-def} the components
of the Fisher information matrix,
\bse
\begin{align}
&\Gamma_{A_0A_0}
\approx \frac{\rho^2}{A_0^2},\quad
\Gamma_{\phi_0\phi_0} \approx \rho^2,
\\[1ex]
&\Gamma_{A_0\phi_0}
\approx \Gamma_{A_0\xi_k} \approx 0,
\quad k=0,\ldots,M,
\\[2ex]
&\Gamma_{\phi_0\xi_k}
\approx \rho^2 \tav{m_k},
\quad k=0,\ldots,M,
\\[2ex]
&\Gamma_{\xi_k\xi_\ell}
\approx \rho^2 \tav{m_k m_\ell},
\quad k,\ell=0,\ldots,M.
\end{align}
\ese

Assuming that the signal \eqref{eq:siglin}
is buried in the stationary and Gaussian noise,
one easily computes its $\F$-statistic,
\be
\label{eq:Fstatl}
\F[x(t);\bxi] \approx \frac{2}{S_0\To}
\left| \int^{\To}_0
x(t)\exp\bigg(-\mi\sum_{k=0}^M\xi_k m_k(t)\bigg)\md t \right|^2,
\ee
where $x(t)$ are the data.

\subsection{False alarm probability}
\label{sSec:false}

Let us calculate the autocovariance function ${\cal C}$
of the $\F$-statistic \eqref{eq:Fstatl}
in the case when data is only noise.
It is defined as
\be
\label{auto3}
{\cal C}(\bxi,\bxi') := \rm{E}_0[\F(\bxi)\F(\bxi')]
- \rm{E}_0[\F(\bxi)]E_0[\F(\bxi')],
\ee
where $\rm{E}_0$ is the expectation value when data is only noise.
We find that $\mathrm{E}_0[\F(\bxi)]=1$ and that
\be
\label{auto5}
{\cal C}(\btau) \approx \tav{\cos\Big(\sum_k \tau_k m_k(t)\Big)}^2
+ \tav{\sin\Big(\sum_k \tau_k m_k(t)\Big)}^2,
\ee
where $\btau:=\bxi-\bxi'$. Thus the autocovariance function
depends only on the difference of the parameters at two points but not on
the parameters themselves.

The autocovariance ${\cal C}$ attains its maximum value
equal 1 when $\btau=\mathbf{0}$.
Let us consider Taylor expansion of ${\cal C}$
around the maximum up to terms quadratic in $\btau$,
\be
\label{TaylorC}
{\cal C}(\btau) \approx 1
+ \sum_k \frac{\partial{\cal C}(\btau)}{\partial\tau_k}
\bigg|_{\btau=\mathbf{0}} \tau_k
+ \frac{1}{2}\sum_{k,\ell}
\frac{\partial^2{\cal C}(\btau)}{\partial \tau_k\partial\tau_l}
\bigg|_{\btau=\mathbf{0}}\tau_k\tau_\ell.
\ee
As ${\cal C}$ attains its maximum
for $\btau=\mathbf{0}$, we have
\be
\label{1stDerC}
\frac{\partial{\cal C}(\btau)}
{\partial\tau_k}\bigg|_{\btau=\mathbf{0}} = 0.
\ee
Let us introduce the symmetric matrix $G$
with elements
\be
\label{eq:G}
G_{k\ell} :=
-\frac{1}{2}\frac{\partial^2 {\cal C}(\btau)}
{\partial\tau_k \partial\tau_\ell}\bigg|_{\btau=\mathbf{0}}.
\ee
One can show that $G = \tilde{\Gamma}$,
where $\tilde{\Gamma}$ is the {\em reduced Fisher matrix}
defined by
\be
\label{redFm-def}
\tilde{\Gamma}_{k\ell}
:=  \tav{m_k m_\ell} - \tav{m_k}\tav{m_\ell}.
\ee
For the linear phase model the components of the reduced Fisher matrix
are constants independent of the values of the parameters.
Making use of Eqs.\ \eqref{1stDerC}--\eqref{redFm-def},
the Taylor expansion \eqref{TaylorC} can be written in the form
\be
\label{TaylorC2}
{\cal C}(\btau) \approx 1
- \sum_{k,\ell}\tilde{\Gamma}_{k\ell}\,\tau_k\,\tau_\ell.
\ee

We define now the {\em correlation hypersurface}
of the statistic $\F$ by the requirement that
the autocovariance $\mathcal{C}$
attains some constant value $\mathcal{C}_0$ on it:
\be
\mathcal{C}(\btau)  = \mathcal{C}_0.
\ee
This equality, by virtue of Eq.\ \eqref{TaylorC2},
can be written as
\be
\label{eq:hype}
\sum_{k,\ell}\tilde{\Gamma}_{k\ell}\,\tau_k\,\tau_\ell = 1 - \mathcal{C}_0.
\ee
Equation (\ref{eq:hype}) defines an $M$-dimensional hyperellipsoid.

The main idea is to divide the space of the phase parameters $\bxi$
into {\em elementary cells} which boundary is determined by Eq.\
\eqref{eq:hype}.
We choose the value $\mathcal{C}_0=1/2$.
We estimate the number $N_\mathrm{c}$ of elementary cells
by dividing the total Euclidean volume $V_{\text{total}}$ of the parameter space
by the Euclidean volume $V_{\text{cell}}$ of the correlation hyperellipsoid,
i.e., we have
\be
\label{NT}
N_\mathrm{c} = \frac{V_\text{total}}{V_\text{cell}},
\ee
where the Euclidean volume of one elementary cell equals
\be
\label{vc}
V_{\text{cell}} = \frac{\pi^{M/2}}{\Gamma(M/2+1)\sqrt{\det G}},
\ee
here $\Gamma$ denotes the Gamma function.

The values of the statistic $\F$ in different cells
can be considered as independent random variables.
We approximate the probability distribution of $\F$
in each cell by the probability distribution
$p_0(\F)$ of $\F$ when the signal is absent.
When the signal is absent,
$2\F$ has a $\chi^2$ distribution with four degrees of freedom.
The false alarm probability $P_F$ for a given cell
is the probability that $\F$ exceeds a certain threshold $\Fo$
when there is no signal;
for $\chi^2$ distribution
with four degrees of freedom we have
\be
\label{PF}
P_F(\Fo) = (1 + \Fo) \exp(-\Fo).
\ee
The probability that $\F$ does not exceed
the threshold $\Fo$ in a given cell is $1-P_F(\Fo)$,
where $P_F(\Fo)$ is given by Eq.\ (\ref{PF}).
Consequently the probability that $\F$
does not exceed the threshold $\Fo$
in \emph{all} the $N_\mathrm{c}$ cells is
$\big(1-P_F(\Fo)\big)^{N_\mathrm{c}}$.
The probability $\alpha$ that $\F$ exceeds $\Fo$ in
\emph{one or more} cell is thus given by
\be
\label{FP}
\alpha = 1 - \big(1 - P_F(\Fo)\big)^{N_\mathrm{c}}.
\ee
This is the desired false alarm probability.
Inverting the formula \eqref{FP} we can calculate the threshold value $\Fo$
corresponding to a chosen false alarm probability $\alpha$.
The expected number of false alarms $N_F$ is given by
\be
\label{NF}
N_F = N_\mathrm{c}\,P_F(\Fo).
\ee

\section{Grid in the parameter space}
\label{Sec:Mac}

In order to search  for a signal in the noise of  the detector we need
to  construct a  grid in  the space  of the  signal's  parameters.  We
define a grid in such a  way that for any possible signal there exists
a grid point in the parameter space such that the expectation value of
the $\F$-statistic  for the parameters  of this grid point  is greater
than a certain value.

In the construction of the grid we employ the approximate linear model
of   the   signal   introduced   in   Sec.\   \ref{sSec:Kro1b}.    The
$\F$-statistic  for this  signal is  given in  Eq.\ \eqref{eq:Fstatl}.
The  expectation  value of  the  $\F$-statistic,  when  the signal  is
present in the data [i.e., when the data $x(t)=n(t)+h(t;\btheta)$,
where $\btheta=(A_0,\phi_0,\bxi)$ collects the signal's parameters], is equal
\be
\mathrm{E}_1\big[\F[n(t)+h(t;\btheta);\bxi']\big]
= 1 + \frac{\rho^2}{2} \mathcal{C}(\btau),
\ee
where  $\rho$ is the  optimal signal-to-noise  ratio computed  in Eq.\
\eqref{snr} and $\mathcal{C}$ is  the autocovariance function given in
Eq.\ \eqref{auto5}; the vector $\btau:=\bxi-\bxi'$, where $\bxi'$ are
the phase parameters of the template.
The function  $\mathcal{C}$ has  the maximum
equal  to  1  for  $\btau=\mathbf{0}$.   The  autocovariance  function
$\mathcal{C}$  is   equal  to  the   square  of  the   match  function
$\mathcal{M}$ originally defined by Owen \cite{Owen96}.

To  construct  the grid  we  first choose  the  minimum  value of  the
correlation   that  we   can  accept.    We  denote   this   value  by
$\mathcal{C}_0$.  (Let us note that $\mathcal{C}_0=\text{MM}^2$, where
MM is the {\em minimal match} introduced by Owen \cite{Owen96}.)  Then we introduce,
as  in Sec.\  \ref{sSec:false},  the correlation  hypersurface of  the
statistic  $\F$  by  the equality  $\mathcal{C}(\btau)=\mathcal{C}_0$,
which, after  making the Taylor  expansion of $\mathcal{C}$ up  to the
second order terms in $\btau$, is described by Eq.\ \eqref{eq:hype}.

\subsection{The covering problem with constraints}
\label{sSec:Maca}

The  problem  of  constructing  a  grid  in  the  parameter  space  is
equivalent to the  so called {\em covering problem} \cite{Con,prix07}:  we want to cover
$(M+1)$-dimensional   parameter  space   with   identical  hyperellipsoids
(\ref{eq:hype}) in such a way that any point of the space belongs to at
least one ellipsoid.
Moreover,  we look  for  an  optimal covering,  i.e.,  the one  having
smallest possible  number of  grid points per  unit volume.   The {\em
  covering  thickness} $\theta$  is  defined  as the  average
number of ellipsoids  that contain a point in  the space.  The optimal
covering would have minimal possible thickness.

Let us  introduce in  the parameter space  the new set  of coordinates
$\mathbf{x}=(x_0,\ldots,x_M)$, defined by the equality
\be
\label{eq:transf}
\btau = \sfM \mathbf{x},
\ee
where the transformation matrix $\sfM$ is given by
\be
\label{eq:mmatr}
\sfM = \sfU_0\,\sfD_0^{-1}.
\ee
Here $\sfD_0$ is the diagonal matrix whose diagonal components are square
roots of eigenvalues of $\tilde{\Gamma}$,  and $\sfU_0$ is a matrix whose
columns  are eigenvectors  of $\tilde{\Gamma}$,  normalized  to unity.
One    can    show   that    $\sfU_0$    is    an   orthogonal    matrix,
$\sfU_0^{-1}=\sfU_0^\mathrm{T}$    (superscript     `T'    denotes    matrix
transposition) and
\begin{equation}
\tilde{\Gamma} = \sfU_0 \, \sfD_0^2 \, \sfU_0^\text{T}.
\end{equation}
Hyperellipsoid  (\ref{eq:hype})   in  coordinates  $\mathbf{x}$   reduces  to  the
$(M+1)$-dimensional  sphere   of  radius  $R:=\sqrt{1-\mathcal{C}_0}$.
Therefore,  the optimal grid  can be  expressed by  means of  a sphere
covering.

In  general,  the  thinnest  possible  coverings  are  known  only  in
dimensions 1 and 2.  In dimensions  up to 5 the thinnest {\em lattice}
coverings  are known [see Eq.\ \eqref{lattice} below for the definition
of a lattice], while in  many higher  dimensions  the thinnest
known coverings  are lattices \cite{Con}.
{From} this point  on we  consider only
lattice coverings.   However, these  general results cannot  be easily
adopted to  our case.   For computational reasons,  we would  like the
nodes of the grid to coincide with Fourier frequencies, so that we can
use the FFT algorithm to calculate the $\F$-satistic efficiently.

Our  grid should  meet  the  following constraint:  one  of its  basis
vectors needs to lie on the  frequency axis and have given length.  In
other words, we look for the  optimal covering with one of the lattice
vectors fixed.  We denote this vector by
\be
\label{eq:avec}
\mathbf{a}_0=(\Delta{}p_0,0,\ldots,0),
\ee
where $\Delta{}p_0$ is the fixed frequency resolution of our procedure.
There is another constraint to be met in an all-sky search:
\be
\label{eq:constraint2}
\mathbf{a}_i=(a_{i0},a_{i1},\ldots{},a_{is},0,0),
\quad i=1,\ldots,s.
\ee
Having  this constraint  satisfied,  we greatly
reduce  the  computational overhead  of  resampling  the  data to  the
barycentric time (see Sec.\ \ref{sSec:Kroa}).

As far as  we know, the general solution to  the covering problem with
constraints  is  not known.   Starting  from  the hypercubic  covering
(i.e., having  all the lattice vectors orthogonal),  a covering satisfying
both constraints can be  constructed, for the signal (\ref{eq:siglin})
with any number of parameters (see \cite{JKbook}).  However, in higher
dimensions  it  may  be   several  times  thicker  than  the  thinnest
unconstrained  lattice known.   An improved  construction  is proposed
here, which takes  as a starting point the  thinnest lattice covering
known in a given dimension, and applies a sequence of modifications to
satisfy the constraints.  We will refer to this lattice as the optimal
covering.

\subsection{Optimal lattice}
\label{sSec:Macc}

A  \emph{lattice}  can  be  conveniently  defined  as a  set  of  all  linear
combinations  of   its  basis  vectors   $\mathbf{a}_i$  with  integer
coefficients:
\be
\label{lattice}
\Lambda=\left\{\sum_ik_i\mathbf{a}_i:
k_i\in\mathbb{Z}\right\}.
\ee
Given lattice  $\Lambda$, its  {\em fundamental parallelotope}  is the
set   of  all points  of   the  form   $\sum_i\theta_i\mathbf{a}_i$,  with
$0\leq\theta_i<1$.  Fundamental  parallelotope is one  example of {\em
  elementary cell}.   The thickness $\theta$ of a  lattice covering is
equal to the  ratio of the volume of one  hyperellipsoid to the volume
of fundamental parallelotope.

For any lattice point $\mathbf{P}_i\in\Lambda$, the {\em Voronoi cell}
around $\mathbf{P}_i$ is defined as
\be
V(\mathbf{P}_i)=\left\{ \btau: C(\btau-\mathbf{P}_i)\geq{}
C(\btau-\mathbf{P}_j)\ \mathrm{for\ all}\ j\neq{}i \right\},
\ee
where  $C(\btau)$  is the  Taylor  expansion  (\ref{TaylorC2}) of  the
autocovariance function.   All Voronoi cells of  any lattice $\Lambda$
are congruent, disjoint, and their  union is the whole space.  Voronoi
cell is  another example  of elementary cell  and is  sometimes called
Wigner-Seitz cell or Brillouin zone.  The Voronoi cell of $\Lambda$ is
inscribed into the correlation ellipsoid (\ref{eq:hype}).

Let $\Lambda$ be any lattice with basis vectors $(\mathbf{a}_0,
\mathbf{a}_1,\dots{})$.  The square of \emph{minimal match} of $\Lambda$ is
\begin{equation}
\label{mm}
\mathrm{MM}^2(\Lambda)
=\inf_{\btau\in{}V(\mathbf{P}_i)}
C(\btau-\mathbf{P}_i),
\end{equation}
where $\mathbf{P}_i$ can be any lattice point.
Let $\bzeta\in V(\mathbf{P}_i)$ be the  value for which the minimum in
(\ref{mm}) is found.  The  function $C(\btau-\mathbf{P}_i)$ has at the
point  $\bzeta$   its  absolute   minimum  inside  the   Voronoi  cell
$V(\mathbf{P}_i)$,  and $\bzeta$ is  a {\em  deep hole}  of $\Lambda$.
Note that  the deep hole  must be one  of the vertices of  the Voronoi
cell.   It  makes  Voronoi  cells especially  useful  for  calculating
minimal match of a given lattice.

We  can now outline  the construction  of an  optimal covering  in the
parameter space.  Given  the value of $C_0$, we  look for the thinnest
possible lattice covering $\Lambda$, satisfying
\begin{equation}
\label{condition}
\text{MM}^2(\Lambda)=C_0.
\end{equation}
As a starting point, we  consider the thinnest lattice covering known.
It is determined by the  number of phase parameters.  For example, the
thinnest covering  of 4-dimensional space  is the so  called Voronoi's
principal lattice  of the first  type $A_4^*$ \cite{Con},  having the
thickness  $\theta_{\mathrm{min}}=1.7655$. The \emph{generator matrix}
(a matrix whose rows are the basis vectors) of this lattice reads
\begin{widetext}
\be
\label{eq:gen}
\sfM_0 = R\sqrt{\frac{5}{2}}
\begin{pmatrix}
\sqrt{2}&0&0&0\\
\frac{1}{\sqrt{2}}&\sqrt{\frac{2}{3}}+\frac{1}{\sqrt{6}}&0&0\\
\frac{1}{\sqrt{2}}&\frac{1}{\sqrt{6}}&
\frac{1}{2\sqrt{3}}+\frac{\sqrt{3}}{2}&0\\
-\frac{1}{5\sqrt{2}}-\frac{2\sqrt{2}}{5}&
\frac{1}{5\sqrt{6}}-\frac{\sqrt{6}}{5}&
-\frac{1}{5\sqrt{3}}-\frac{\sqrt{3}}{10}&
-\frac{1}{2\sqrt{5}}
\end{pmatrix},
\ee
\end{widetext}
where $R$ is the covering radius.

Let  $\mathbf{l}$  be  any   lattice  vector  of  $A_4^*$,  such  that
$|\mathbf{l}|\geq{}|\sfM^{-1}\mathbf{a}_0|$, where $\sfM$ and $\mathbf{a}_0$
are  given  by  Eqs.\ (\ref{eq:mmatr})  and  (\ref{eq:avec}),  respectively.
The vector $\mathbf{l}$ is  a linear combination  of rows of  (\ref{eq:gen}).
In
order   to  construct  an   optimal  lattice,   satisfying  constraint
(\ref{eq:avec}),  we  perform   two  scaling  operations  on  $A_4^*$:
shrinking   in   the  direction   of   $\mathbf{l}$   by  the   factor
$\mu=|\sfM^{-1}\mathbf{a}_0|/|\mathbf{l}|\leq{}1$  and  expanding in  all
directions perpendicular to $\mathbf{l}$ by the factor $\nu\geq{}1$.
The generator matrix of the shrunk lattice is
\be
\sfM_1 = \sfM_0 \,\sfO_1 \,\mathrm{diag}(\mu,1,1,1) \,\sfO_1^{-1},
\ee
where  $\sfO_1$  is  an  orthogonal transformation  defined  by
the condition that $\sfO_1(1,0,0,0)=\mathbf{l}/|\mathbf{l}|$.

The  expansion factor  $\nu$  is  defined by  the  condition that  the
covering  radius $R$ remains  unchanged on  scaling.  In  general, its
value can be determined numerically using the following iteration:
\be
\sfM_{i+1} = \sfM_i \,\sfO_1 \,\mathrm{diag}(1,\nu_i,\nu_i,\nu_i) \,\sfO_1^{-1},
\quad\text{for  $i=1,2,\dots$},
\ee
where  $\nu_i=R/(R_i\sin{\phi_i})$, $R_i$  is the
covering radius  of the lattice given by  $\sfM_i$, and $\phi_i\leq\pi/2$
is  the largest  angle that  a half-line  starting at  the  origin and
containing  a  deep hole  of  $\sfM_i$ can  form  with  the direction  of
$\mathbf{l}$.
The  above  procedure  converges  after several  iterations,  and  the
expansion factor is formally $\nu=\prod_{i=1}^{\infty}\nu_i$.

After scaling, the lattice thickens by a factor $1/(\mu\nu^3)\geq{}1$,
depending on the choice of $\mathbf{l}$. Note that $\mu\nu^3\to{}0$ as
the length of $\mathbf{l}$  increases.  By enumerating the lattice vectors
of $A_4^*$ in order of  increasing magnitude, one can find the optimal
$\mathbf{l}$, such that $1/(\mu\nu^3)$ is minimal.

The generator matrix of an optimal lattice, satisfying the constraint
(\ref{eq:avec}), is
\begin{equation}
\label{eq:opt}
\sfM_{\mathrm{opt}} = \sfM_0 \,\sfO_1 \,\sfS \,\sfO_2^{-1} \,\sfM^\mathrm{T},
\end{equation}
where  $\sfO_2$  is  an  orthogonal  transformation  satisfying
$\sfO_2(1,0,0,0)=\mathbf{e}_0/|\mathbf{e}_0|$
with $\mathbf{e}_0=\sfM^{-1}\mathbf{a}_0$,
$\sfS$ is a diagonal  matrix with elements  $(\mu,\nu,\nu,\nu)$.
The lattice vector $\mathbf{l}$ is  chosen in such  a way that $\det\sfS$ is
maximal.
The thickness of this lattice is
\begin{equation}
\theta = \frac{\theta_{\mathrm{min}}}{\det\sfS}.
\end{equation}

For the  case of the  Virgo antenna, observational time  $T_0=2$ days,
frequency  750  Hz,  effective  bandwidth  1 Hz,  and  $C_0=3/4$,  the
resulting   lattice   is   thicker   than  $A_4^*$   only   by   20\%,
$\theta\simeq{}2.1$, reducing  the number  of templates to  roughly 10
millions (the corresponding hypercubic  lattice would have 25 millions
of templates).

The  constraint   (\ref{eq:constraint2})  can  be   satisfied  without
increasing   the  lattice  thickness   by  a   4-dimensional  rotation
$\sfO_3$ such that $\sfO_3\mathbf{e}_0=\mathbf{e}_0$ and
the  vector  $\sfM\sfO_3\sfO_2\sfO_1^{-1}\mathbf{f}$
is orthogonal  to the $(\alpha_1,\alpha_2)$  plane, where $\mathbf{f}$
is  a lattice  vector of  $A_4^*$, $\mathbf{f}\nparallel{}\mathbf{l}$.
The generator matrix of a lattice satisfying both constraints is now
\be
\label{eq:opt2}
\sfM_{\mathrm{opt}} = \sfM_0 \,\sfO_1 \,\sfS \,\sfO_2^{-1} \,\sfO_3^{-1}\,
\sfM^\mathrm{T}.
\ee

\subsection{Two-dimensional example}
\label{sSec:Macd}

\begin{figure}
\scalebox{0.9}{\includegraphics{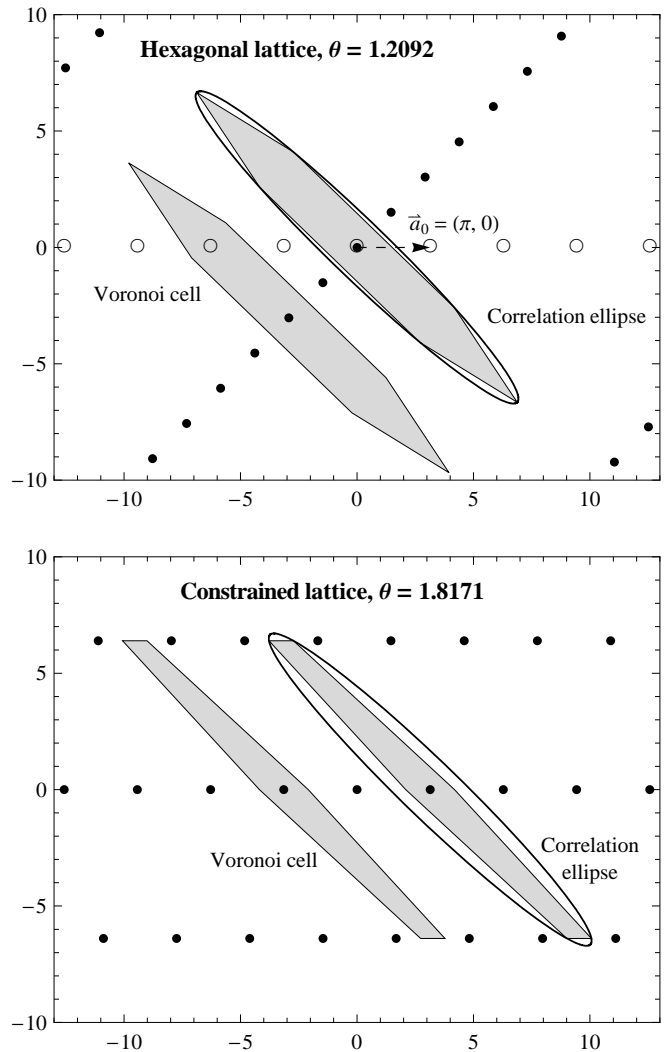}}
\caption{\label{fig:lattice2d}
\textit{Up:} Hexagonal covering generated by
(\ref{eq:gen2hex}), with  $\mathcal{C}_0=3/4$  and $R=1/2$  (dark
points).   Fourier frequencies  are represented  by  open circles.
\textit{Down:} Constrained lattice, satisfying the condition (\ref{eq:constr2d}).
Any point in the parameter space belongs on average to 1.8171 ellipses.}
\end{figure}

Let us  explain the  construction of an  optimal lattice on  a simple,
two-dimensional  example. We consider here the signal
\be
\label{2dimsignal}
h(t;A_0,\phi_0,\bar{\omega}_0,\bar{\omega}_1)
= A_0 \cos\bigg(\bar{\omega}_0 \frac{t}{\To}
+ \bar{\omega}_1 \Big(\frac{t}{\To}\Big)^2 + \phi_0\bigg),
\ee
where $\To$ is the observation time
and the observation interval is $\left\langle0,\To\right\rangle$.
The  phase  of the  signal \eqref{2dimsignal} depends on  two  dimensionless
parameters $\bar{\omega}_0$, $\bar{\omega}_1$ (and on the initial phase parameter $\phi_0$).
The \emph{reduced} Fisher matrix for this signal reads
\begin{equation}
  \label{eq:gamma2d}
  \tilde{\Gamma}=
\begin{pmatrix}
\frac{1}{12}&\frac{1}{12}\\[1ex]
\frac{1}{12}&\frac{4}{45}
\end{pmatrix}.
\end{equation}
The correlation hypersurface (\ref{eq:hype}) is now an ellipse:
\be
  \label{eq:ellipse}
  \frac{(\tau_1+\tau_2)^2}{12}+\frac{\tau_2^2}{180}=1-\mathcal{C}_0.
\ee
Correlation   ellipse    is   shown   on   both    parts   of Fig.\
\ref{fig:lattice2d}.     Frequency    resolution   when    calculating
$\F$-statistic is  now $\pi$, therefore  we require the  first basis
vector of a lattice to be equal to
\begin{equation}
  \label{eq:constr2d}
  \mathbf{a}_0=(\pi,0),
\end{equation}
in order to satisfy the constraint (\ref{eq:avec}).

The  thinnest possible  covering in  two dimensions  is  the hexagonal
lattice, $A_2^*$.  The  Voronoi cell of $A_2^*$ is  a regular hexagon,
and           the            covering           thickness           is
$\theta_{\text{hex}}=2\pi/(3\sqrt{3})\cong{}1.2092$.       The generator
matrix of $A_2^*$ is
\begin{equation}
  \label{eq:A2gen}
  \sfM_0=
  R\sqrt{3}
    \begin{pmatrix}
      1&0\\
      \frac{1}{2}&\frac{\sqrt{3}}{2}
    \end{pmatrix},
\end{equation}
where $R=\sqrt{1-\mathcal{C}_0}$ is the  covering radius.  In order to
cover the parameter space with ellipses (\ref{eq:ellipse}), we have to
transform the hexagonal lattice  to the $\btau\equiv(\tau_1,\tau_2)$ coordinates, as given
by (\ref{eq:transf}).   The generator matrix of  an optimal, hexagonal
covering in $\btau$ coordinates is
\begin{equation}
  \label{eq:gen2hex}
  \sfM_{\mathrm{hex}} = \sfM_0 \, \sfM^{\mathrm{T}}.
\end{equation}
The  resulting  lattice   is  shown  in  the  upper   part  of  Fig.\
\ref{fig:lattice2d}, along  with the correlation  ellipse. The Voronoi
cell is a hexagon inscribed  into the correlation ellipse (any lattice
$\Lambda$  constitutes  a covering  if  and  only  if the  correlation
hypersurface completely includes its Voronoi cell).  Note that lattice
points do  not coincide with Fourier frequencies,  represented by open circles.

The requirement  (\ref{eq:constr2d}) can be satisfied by  the means of
transformation  (\ref{eq:opt}),  at  the  cost of  increasing  lattice
thickness.  We  find that $|\sfM^{-1}\mathbf{a}_0|=\pi/(2\sqrt{3})$.  For
$\mathcal{C}_0=3/4$ (then $R=1/2$),  the  best  choice of  $\mathbf{l}$  is
$\mathbf{l}=\sqrt{3}/4(3,\sqrt{3})$.  The hexagonal lattice in coordinates
$x_i$   is    shrunk   in    the   direction   of    $\mathbf{l}$   by the
factor
$\mu=|\sfM^{-1}\mathbf{a}_0|/|\mathbf{l}|=\pi/(3\sqrt{3})$,  then expanded  by
the factor
$\nu=\sqrt{108-\pi^2}/9$, and finally rotated in such a  way that $\mathbf{l}$
coincides with $\sfM^{-1}\mathbf{a}_0$.

The  lattice  such obtained is   shown  in  the   lower  part   of  Fig.\
\ref{fig:lattice2d}. It        has        the        thickness
$\theta=\theta_{\mathrm{hex}}/(\mu\nu^3)=18/\sqrt{108-\pi^2}\cong1.8171$. Note
that the Voronoi cell has changed. It  covers more than half of the area of the
correlation ellipse and is inscribed in it.

\section{Position and velocity of a~detector with respect to the solar system
barycenter}
\label{Sec:Kaz}

In this section we describe algorithms and procedures
needed to compute, for a given time and a geographical location of a detector,
the corresponding vectors of position and velocity
of the detector with respect to the Earth barycenter and of this
barycenter with respect to the solar system barycenter (SSB).
The sum of these vectors represents
the position and velocity of the detector referred to the SSB.

We have incorporated the algorithms and procedures
described in the present section
into a set of \textsc{fortran} subroutines
which we have called the \texttt{Top2Bary} package.

\subsection{Overview of transformations involved}
\label{sSec:Kaza}

\subsubsection{Reference systems and frames\protect\footnote{
We employ here the recent recommendations of the IAU \cite{Kaplan2005}.}}

Reference data for positional astronomy,
such as the data in barycentric planetary epheme\-ri\-des,
are now specified within the International Celestial Reference System (ICRS).
The ICRS is a coordinate system whose
origin is at the SSB and whose axis directions are effectively
defined by the adopted coordinates of 212 extragalactic radio sources which are
assumed to have no observable intrinsic angular motions. Thus, the ICRS is
a ``space-fixed'' system (more precisely, a kinematically non-rotating system)
without an associated epoch. However, the ICRS closely matches the conventional
dynamical system defined by the Earth's mean equator and equinox of J2000.0;
the alignment difference is at the 0.02 arcsecond level, negligible for many applications.
The list of radio source positions that define ICRS for practical purposes is
called the International Celestial Reference Frame (ICRF).

The position and velocity 3-vectors taken from the JPL DE405/LE405
ephe\-meris are in equatorial rectangular coordinates referred
to the SSB. The reference frame for the DE405
is the ICRF; the alignment onto this frame, and therefore onto the ICRS,
has an estimated accuracy of a few milliarcseconds,
at least for the inner-planet data.

The DE405 was developed using T$_\mathrm{eph}$,
a barycentric coordinate time \cite{S98}. T$_\mathrm{eph}$ is rigorously
equivalent to Barycentric Coordinate Time (TCB) in a mathematical
sense, differing only in rate: the rate of T$_\mathrm{eph}$ matches the average
rate of TT (Terrestrial Time, or TDT), while the rate of TCB is defined
by the SI system. The IAU time scale, Barycentric Dynamical Time
(TDB), often (but erroneously) considered to be the same as T$_\mathrm{eph}$,
is a quantity that cannot be physically realized, due to its flawed
definition. So, in fact, the use of the name TDB actually refers to
quantities based on or created with T$_\mathrm{eph}$ (because of this,
the IAU Working Group on Nomenclature for Fundamental Astronomy
has recommended changing the definition of TDB to be consistent
with that of T$_\mathrm{eph}$).
Astronomical constants obtained from ephemerides based on T$_\mathrm{eph}$
(or TDB) are not in the SI system of units and must therefore be scaled
for use with TCB or other SI-based time scales.

The epoch J2000.0 is the epoch of 2000 January 1, 12$^\mathrm{h}$~TT
(JD 2451545.0 TT) at the geocenter (``J2000.0 system'' is shorthand for the celestial
reference system defined by the mean dynamical equator and equinox of J2000.0).
The coordinate system defined by the ``equator and equinox
of J2000.0", can be thought of as either barycentric or geocentric.

It is also worth noting that the recent IAU resolutions do not describe
the proper reference system of the observer---the local, or topocentric,
system in which most measurements are actually taken. The resolutions
as adopted apply specifically to Einstein's theory of gravity,
i.e., the general theory of relativity.

\subsubsection{Time scales}

It is assumed that the time, associated with observational data we are
dealing with, is the Coordinated Universal Time (UTC)
as disseminated by international time services. The UTC scale since 1972
is essentially uniform, except for occasional 1 second steps (leap seconds)
introduced internationally to compensate for the variable
Earth rotation. By 2010 there were 24 leap seconds. Earlier,
1961 to 1971, the adjustments were continuous.
UTC devoid of these adjustments is called
the International Atomic Time, TAI.
Addition of the difference TAI$-$UTC
to UTC converts it to TAI which is uniform.\footnote{
The TAI$-$UTC differences are available in a tabular form
at \url{hpiers.obspm.fr/eoppc/bul/bulc/UTC-TAI.history}.
The function \texttt{tai\_ut} of our \texttt{Top2Bary} package
reads similar file and returns the difference
calculated for given UTC Julian Date.}
TAI in turn differs from the Terrestrial (Dynamical) Time, TDT or TT
(normally used to describe celestial phenomena by astronomers),
only by a constant term:
$$
\text{TDT} = \text{TAI} + 32.184\,\mathrm{s}.
$$

As already explained, the time argument of the JPL positions and velocities of
celestial objects is in principle T$_\mathrm{eph}$
or the barycentric coordinate time.
This time scale in practice can be equated
with the Barycentric Dynamical Time, TDB
(in spite of the noted inadequacy in its definition).
The TDB differs from the TDT only by small periodic terms, which to
sufficient accuracy are usually simplified to only two largest terms:
$$
\text{TDB} - \text{TDT}
= \big(0.001658\,\sin g + 0.000014\,\sin 2g \big)\,\mathrm{s},
$$
where $g=\big(357.53+0.9856003$ (JD$-$2451545.0)\big) degrees
and JD is the Julian Date equal to MJD + 2400000.5,
MJD being the Modified Julian Date.
So essentially, the two scales do not have to be distinguished
and for many practical purposes can be assumed equal,
TDB$=$TDT.\footnote{
In the \texttt{Top2Bary} package we do distinguish them,
but the mentioned possibility can be made effective
by setting the \texttt{iTDB} option,
in the \texttt{useTop2B.cfg} configuration file, to 0.}

One can relate given UTC to barycentric positions
of all the major celestial bodies of the solar system
as obtainable from the JPL ephemeris.\footnote{These relations are
incorporated into the \texttt{Top2Bpv} subroutine, which also calls
the \texttt{tai\_ut} function that reads a file which contains details
of the UTC adjustments.}
However, to relate the position of a point on the Earth to the geocentric
ICRF (i.e., the same frame as the barycentric ICRF except for the origin
of axes which is now at the barycenter of the Earth)
one has to use yet another time scale---the rotational time scale UT1,
which is nonuniform and is determined from astronomical observations.
The difference UT1$-$UTC, the value of which is presently maintained by
international services within $\pm0.90$~s, is taken from the International
Earth Rotation Service (IERS) tabulations
of daily values publicly available as \texttt{eopc04.yy} files.\footnote{
Here \texttt{yy} stands for a two-digit year number
(e.g., \texttt{99} for 1999, and \texttt{06} for 2006).
The \texttt{eopc04.yy} files are available at the IERS Internet site
\url{www.iers.org/MainDisp.csl?pid=36-25788&prodid=22}.
In the \texttt{Top2Bary} package
the UTC to UT1 conversion takes place
in the \texttt{sitePV} subroutine, which calls a polar motion routine,
\texttt{polmot}. The latter returns the UT1$-$UTC difference
interpolated (between two nearest midnight values read from appropriate
\texttt{eopc04} file) to the UTC given,
along with other parameters of Earth axis motion
(the terrestrial and celestial pole offsets) similarly interpolated.}
The UT1 time can be readily converted
to the Earth rotational angle or the sidereal time.\footnote{
In the \texttt{Top2Bary} package
this conversion is performed in the \texttt{sid} function, which is
a function of UT1 and the location geographical longitude corrected for
the polar motion.}

\subsubsection{Location coordinates and velocities with respect
 to the Earth bary\-center}

To be able to express coordinates of a point on the Earth surface
in the Earth centered ICRF,
it is necessary to know orientation of the Earth in space.
The primary effects that should be taken into account are:
diurnal (variable) rotation, precession and nutation of the Earth rotational
axis, and polar motion.

The precession is accounted for by applying
standard astronomical theories. We use the new IAU theory \cite{Lieske1979}.
The nutation also could be computed basing on a theory, but the DE405 has
it in the numerical form, so we just read the nutation angles, $\Delta\psi$
and $\Delta\varepsilon$. These nutation angles are not the same as defined
in the newest IAU nutation theory, so when highest
precision is required the celestial pole offsets, d$\psi$ and d$\varepsilon$,
must also be added to these angles (in the past the magnitude of these
offsets remained below 0.1 arcsecond).
For past years, since 1962, these two offsets are included
in the already mentioned \texttt{eopc04.yy} files.\footnote{
Our package normally adds these offsets, but the user may change this option
by setting  the \texttt{NutSid} parameter in the configuration file
\texttt{useTop2B.cfg}. Note, however, that doing so he will affect not only
nutational transformations but also computing of the equation of equinoxes
which enters formulas for the sidereal time
and thus affects the rotation angle of the Earth.}

The remaining two effects, the Earth variable rotation and polar motion,
are unpredictable for a remoter future, so also observational data must
be used. The data necessary for reduction are taken from the
\texttt{eopc04.yy} files as well.
The polar motion can be taken into account by modifying the conventional
geographical coordinates of a point on the Earth [see Eqs.\ (5.1) in
\cite{ABJK02}] or modifying the rectangular coordinates corresponding
to these conventional geographical coordinates:
\bse
\begin{align}
x_\circ &= r_\circ \cos\lambda_\circ,
\\[1ex]
 y_\circ &= r_\circ \sin\lambda_\circ,
\\[1ex]
z_\circ &= b \sin\Psi + h \sin\phi_\circ,
\end{align}
\ese
where $\phi_\circ$ is the conventional geographical latitude,
$\lambda_\circ$ --- the conventional longitude,
$h$ --- the height above the Earth ellipsoid,
$\Psi=\arctan(b \tan\phi_\circ /a)$ --- the reduced latitude,
$r_\circ=a \cos\Psi + h \cos\phi_\circ$ is the equatorial
component of the radius vector, and $a=6378.140$ km and
$b=a(1-1/f)$ are the semiaxes of the ellipsoid
(the flattening $f$ is taken equal to 1/0.00335281,
which is the NOVAS value\footnote{
The user may change the value of the flattening $f$
to the IAU value of 298.257,
by changing the \texttt{NovF} parameter in the configuration file.}).

In this version we have adopted the second possibility (i.e.\ correcting
of the Cartesian coordinates) using the following relations:
\bse
\begin{align}
x &= x_\circ - P_x z_\circ,
\\[1ex]
y &= y_\circ + P_y z_\circ,
\\[1ex]
z &= z_\circ + P_x x_\circ - P_y y_\circ,
\end{align}
\ese
where  $P_x$ and $P_y$ are the IERS coordinates of the pole,
with respect to the Conventional International Origin
(to which the `conventional' geographical coordinates refer), converted to radians.

These $(x,y,z)$ coordinates are expressed in the terrestrial reference
frame with the $x$ axis directed toward the Greenwich meridian. To relate
them to the celestial frame, the ICRF, the rotational angle of the Earth
must be taken into account. This is done through conversion of UTC to UT1
(as described in the preceding subsection). The UT1 time serves to calculate
the apparent (or true) local sidereal time $\theta$
(returned by the \texttt{sid} function),
which includes the nutational component
(nutation in longitude corrected for the corresponding IERS celestial pole
offset
also taken from the IERS \texttt{eopc04} files).
The apparent local time is advanced with respect to
the mean Greenwich sidereal time by the true location longitude
[calculated as $\lambda=\arctan(y/x)$
with proper choice of one of the four quadrants]
and the mentioned nutational component.
This component is equal to the so called
{\em equation of equinoxes} $(\Delta\psi+\md\psi)\cos\varepsilon$
plus a very small correction to this equation
(which amounts to less than $0.003''$ and depends on the mean
longitude of the ascending node of the Moon).\footnote{
Our configuration file allows the user to neglect this small correction
or even use the mean sidereal time instead.}

So computed local sidereal time $\theta$ is finally used
to find rectangular coordinates of the location in the geocentric ICRF:
\bse
\begin{align}
X &= r_e\cos\theta,
\\[1ex]
Y &= r_e\sin\theta,
\\[1ex]
Z &= z,
\end{align}
\ese
where $r_e = \sqrt{x^2 + y^2}$ is the equatorial component.

At this point the location velocity due to Earth rotation can be
computed. The location motion relative to
the Earth barycenter is represented by a vector of constant length
[in principle, $v_\circ=2\pi r_e/(\text{sidereal day})=\Omega_\text{r}r_e$,
where $\Omega_\text{r}$ is the Earth angular rotation speed]
and directed always towards the east in the topocentric reference frame.
This diurnal velocity vector has the following Cartesian components:
\bse
\begin{align}
V_x &= v_\circ\cos(\theta+\pi/2) = -v_\circ\sin\theta,
\\[1ex]
V_y &= v_\circ\sin(\theta+\pi/2) = +v_\circ\cos\theta,
\\[1ex]
V_z &= 0,
\end{align}
\ese
where the numerical value of $v_\circ$ is calculated
with the NOVAS constant of the Earth angular rotation speed,
$\Omega_\text{r}=2\pi/$(sidereal day in TAI seconds)$= 7.2921151467\times10^{-5}$
rad/s.\footnote{ This constant corresponds to \texttt{NOmega} (a parameter
listed
in the configuration file) set to 1, and can be changed to the IERS
value of $7.292115\times10^{-5}$ (\texttt{NOmega} = 0) or to
$2\pi/(24\times 3600)\times1.002737909350795 = 7.2921158553\times10^{-5}$
(\texttt{NOmega} = 2). All the above conversions
are done in the \texttt{sitePV} subroutine.}

Since our approach is essentially classical, these Cartesian coordinates
($X,Y,Z$) and velocities ($V_x,V_y,V_z$)
are naturally referred to the frame of equator and equinox of date.
Therefore they are further nutated and precessed (in this order) back to
the standard epoch J2000.0.\footnote{This is performed in the \texttt{Top2Bpv}
subroutine by calling the \texttt{RemNut} and \texttt{prexyz} procedures,
separately for the position vector and velocity vector.}

\subsubsection{Barycentric coordinates of the Earth (JPL ephemeris)}

For computing the coordinates of the Earth barycenter, relative to the SSB,
use is made of the fundamental solar system
ephemerides from the Jet Propulsion Laboratory (JPL).
The latest JPL Planetary and Lunar Ephemerides, DE405/LE405 or
just DE405, were created in 1997
and are described in detail in Ref.\ \cite{S98}.\footnote{
They are available via the Internet
(\url{ssd.jpl.nasa.gov/?planet_eph_export})
or on CDrom (from the publisher: Willmann-Bell, Inc.;
\url{www.willbell.com/software/jpl.htm}).}
The DE405 ephemeris is based upon the ICRF.\footnote{
An earlier popular ephemeris DE200, which has been the basis
of the {\em Astronomical Almanac} since 1984,
is within 0.01 arcseconds of the ICRF.}
It constitutes of a set of Chebyshev polynomials fit with
full precision to a numerical integration over 1600 AD to 2200 AD.
The JPL package allows to obtain the rectangular coordinates
of the Sun, Moon, and nine major planets anywhere between JED
(i.e., Julian Ephemeris Date) 2305424.50 (1599 Dec 09)
and JED 2525008.50 (2201 Feb 20).
Besides coordinates, it includes nutations and librations.\footnote{
Our routines do make use of the JPL nutation in longitude
and in obliquity after correction for (addition of) the IERS celestial pole
offsets.
We have used only a 21-year (1990 to 2010) subset of the original ephemeris.}

The ephemeris gives separately the position and velocity of the
Earth-Moon barycenter and the Moon's position and velocity relative to this
barycenter.
The Earth position and velocity vectors (relative to the Earth-Moon
barycenter) are thus calculated as a fraction
(involving the masses of the two bodies) of the Moon's vectors and opposite
to them. For example, the $x$ coordinate of the Earth barycenter with
respect to the SSB is obtained as:
$$
x_\text{E} = x_\text{EM} - x_\text{M}/82.30056,
$$
where `EM' and `M' subscripts refer to the Earth-Moon barycenter
and the Moon, respectively, and the numerical denominator is equal to
the Earth-plus-Moon to Moon ratio of masses taken from the JPL ephemeris.
Similar expressions pertain to the $y_\text{E}$ and $z_\text{E}$ coordinates.
Since the DE405/LE405 coordinates are given in the J2000.0 reference frame,
the position and velocity of the Earth barycenter so obtained
need not be nutated nor precessed.\footnote{
The position and velocity vectors of the Earth are computed
by calling the \texttt{EarthPV} subroutine, which in turn calls
the original JPL \texttt{STATE} routine slightly modified for our needs.
This subroutine reads the JPL planetary ephemeris file named
\texttt{DE405'90.'10} and interpolates the data to the specified epoch.}

Finally, the velocity vector of the motion of the Sun towards its apex
(with the speed of 20 km/s) can be optionally added
to the Earth barycenter velocity. The direction of solar apex is assumed
at 18$^\text{h}$ in right ascension and 30$^\circ$ in declination
in the frame of equator and equinox of J1900.0.
Therefore this direction has to be precessed
from that epoch to J2000.0.\footnote{
Since sky positions in astronomical catalogues
are not corrected for the apex motion, this component
is actually only optionally included in the \texttt{Top2Bary} package
(corresponding \texttt{iSunV} option is set to 0 in the code). If desired,
the \texttt{iSunV} option can be changed by the user just by setting
a nonzero value in the configuration file, \texttt{useTop2B.cfg}. In this
case the velocity vector towards the apex is added to the Earth velocity
vector (position remaining unaffected).}

\subsection{Structure of the {\tt Top2Bary} module}
\label{sSec:Kazb}

\begin{figure}
\scalebox{0.5}{\includegraphics{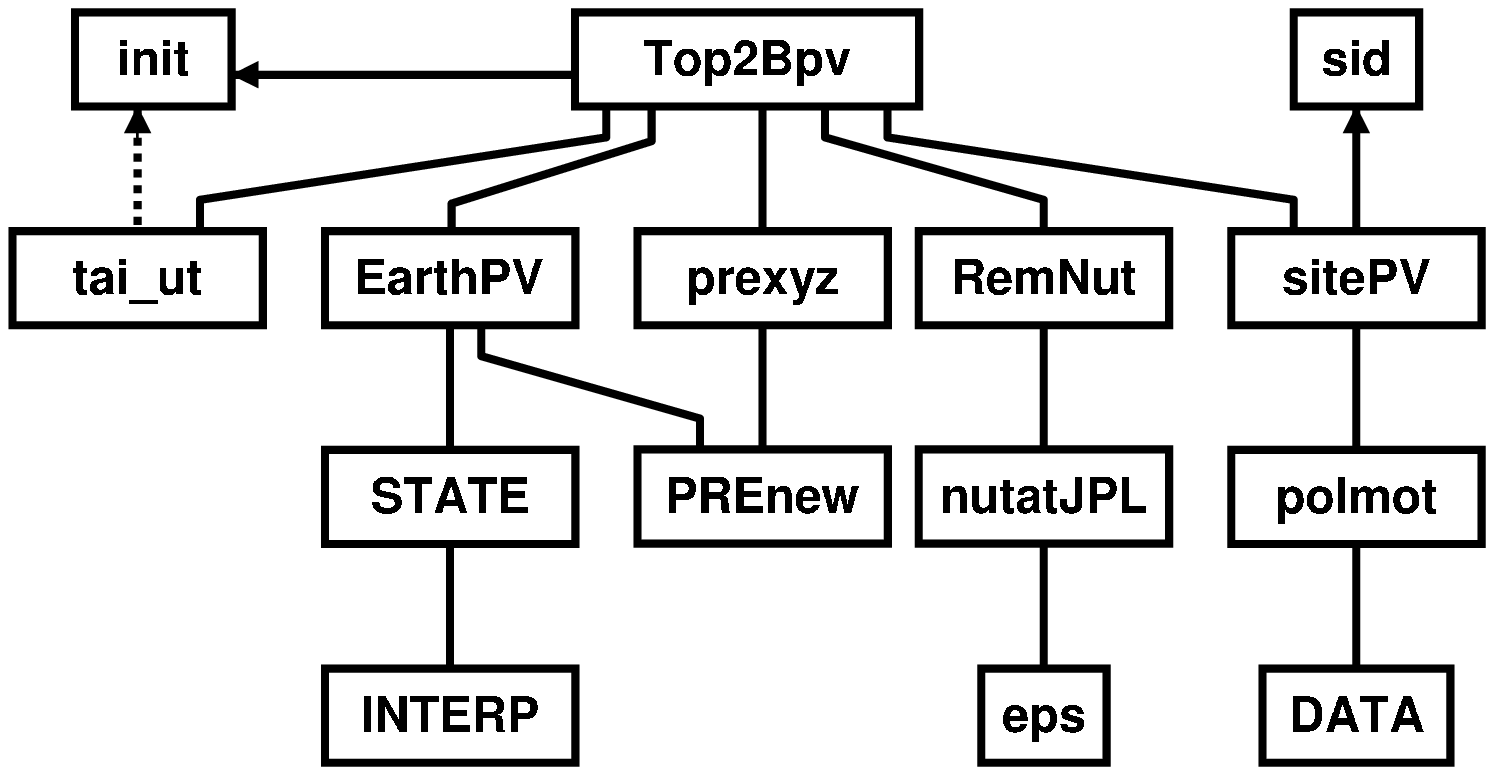}}
\caption{\label{Top2Bary}
Block diagram of the {\tt Top2Bary} module.}
\end{figure}

The above described algorithms and procedures
were implemented in a module named \texttt{Top2Bary},
consisting of about 900 lines of \textsc{fortran} code.
The overall structure of the module is shown in Fig.\ \ref{Top2Bary}.
The {\tt Top2Bary} module contains the following subprograms:
\texttt{Top2Bpv}---main subroutine (to be called by user);
\texttt{init}---sets some constants and parameters;
\texttt{sitePV}---computes location geocentric vectors;
\texttt{EarthPV}---computes Earth barycentric vectors;
\texttt{STATE}---modified JPL \texttt{STATE} that reads DE405;
\texttt{INTERP}---original JPL subprogram;
\texttt{prexyz}---precesses rectangular coordinates using \texttt{PREnew};
\texttt{PREnew}---computes general precession
(within the new IAU theory\footnote{The once adequate word 'new'
is misleading in view of recent revolutionary changes of concepts
since by this name really referred is here that old IAU 1976 theory
\cite{Lieske1979}.});
\texttt{RemNut}---eliminates nutation from a vector;
\texttt{nutatJPL}---returns JPL nutation angles and mean obliquity;
\texttt{eps}---calculates mean obliquity;
\texttt{sid}---computes local sidereal time (mean or apparent);
\texttt{tai\_ut}---calculates the difference of TAI$-$UTC;
\texttt{polmot}---interpolates IERS UT1$-$UTC and pole offsets;
\texttt{DATA}---finds Gregorian or Julian calendar date from JD.

The module expects the presence of the following data files:
\texttt{DE405'90.'10}---JPL DE405 ephemeris binary file
spanning the years 1990 to 2010;
\texttt{tai-utc.dat}---table of the TAI$-$UTC differences (ASCII);
\texttt{eopc04.yy}---IERS files, one per \texttt{yy}-year, for desired years
(ASCII);
\texttt{useTop2B.cfg}---optional configuration file (ASCII).\footnote{
Except for the last file which is looked for in the current directory,
all the other files must be placed in the \texttt{/Top2Bary/EphData}
subdirectory.
All the ASCII files \textit{must} be formatted in Windows style
(i.e., the lines must be terminated with CR/LF) and not UNIX style (with LFs
only).
This is important for a user who updates or modifies existing files
or downloads new \texttt{eopc04.yy}\ files from the IERS site,
where they are stored in the UNIX format.}

\section{Search algorithm}
\label{Sec:Kro}

In the case of all-sky searches for gravitational-wave signals
from rotating neutron stars the parameter space is very large
and it is important to calculate the $\F$-statistic as efficiently as possible.

\subsection{Resampling}
\label{sSec:Kroa}

The detection statistic $\F$ of Eq.\ (\ref{OS}) involves integrals given by Eqs.\ (\ref{Fab}).
Let us consider the integral (\ref{eq:Fa}) [the same arguments will apply to the integral (\ref{eq:Fb})].
The phase $\phi(t)$ [see Eq.\ (\ref{pha3})] can be written as
\be
\phi(t) = \omega_0 [t + \phim(t)] + \phis(t),
\ee
where
\bse
\begin{align}
\label{pham}
\phim(t) &:= \frac{{\bf n}_0\cdot{\bf r}_{\rm d}(t)}{c},
\\
\label{phas}
\phis(t) &:= \sum_{k=1}^{s}\omega_k \frac{t^{k+1}}{(k+1)!}
+ \frac{{\bf n}_0\cdot{\bf r}_{\rm d}(t)}{c} \sum_{k=1}^{s}
\omega_k\frac{t^k}{k!}.
\end{align}
\ese
The functions $\phim(t)$ and $\phis(t)$ do not depend on frequency $\omega_0$.
We can write the integral \eqref{eq:Fa} as
\be
\label{eq:Far}
F_a = \int_0^\To x(t)\,a(t)\,e^{-\mi\phis(t)}
\exp\big\{-\mi\omega_0[t + \phim(t)]\big\} \, \md t.
\ee
Next we introduce a new time variable $\tb$,
so called {\em barycentric time} \cite{S91,JKS98},
\be
\label{eq:Bt}
\tb(t) := t + \phim(t).
\ee
In this new time coordinate the integral (\ref{eq:Far})
is approximately given by (see Ref.\ \cite{JKS98}, Sec.\ III~D)
\be
\label{ia}
F_a \cong \int_0^\To x[t(\tb)] a[t(\tb)]
e^{-\mi\phis[t(\tb)]} e^{-\mi\omega_0\tb}\,\md\tb.
\ee
This integral is a Fourier transform of the data $x[t(\tb)]$ multiplied by the
function $a[t(\tb)]\exp[-\mi\phis[t(\tb)]]$.
For discrete data $x(t)$ the integral (\ref{ia}) can be converted
to a discrete Fourier transform which can be evaluated by the FFT algorithm.

\begin{figure}
\scalebox{0.50}{\includegraphics{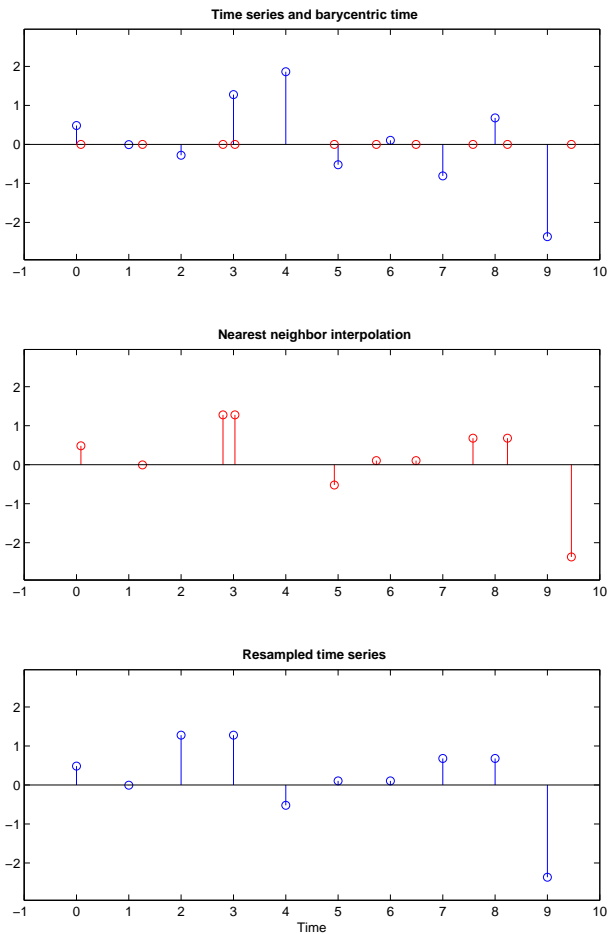}}
\caption{\label{fig:view_r}
Illustration of the nearest neighbor resampling method.
The top panel shows the uniformly sampled original data
and the points of the barycentric time.
The middle panel shows the interpolations of the original time series
at the points of the barycentric time,
obtained by the nearest neighbor method.
The bottom panel shows the resampled time series
which is the uniformly sampled barycentric time series form the middle panel.}
\end{figure}

Thus to convert the integral (\ref{eq:Far}) into a Fourier transform
we need to resample the function $x(t)\,a(t)\,e^{-\mi\phis(t)}$ according to Eq.\ (\ref{eq:Bt}).
We consider two numerical interpolation methods
in order to obtain the resampled function.
The first method is the {\em nearest neighbor interpolation}
also called the {\em stroboscopic resampling}. We assume that the original
data is a time series $x_k$ ($k=1,\ldots,N$), sampled at uniform intervals.
In this method we obtain the value of the time series $x_k$
at barycentric time $\tb$ by taking the value $y_{k_0}$
such that $k_0$ is the nearest integer to $\tb$.
We have illustrated the method in Fig.\ \ref{fig:view_r}.

The second method has two steps. The first step consists of obtaining
a more finely sampled time series and the second step consists of interpolating
the
upsampled time series to the barycentric time using splines \cite{boor-78}.
To perform the first step we use an interpolation method based on the Fourier
transform.
We take the Fourier transform of the original time series,
pad the Fourier transform series with an appropriate amount of zeros
and then transform it back to the time domain by inverse Fourier transform.
The Fourier transforms are performed using the FFT algorithm.
We thus obtain an interpolated time series with points inserted between the
original points.
If we have a time series with $N$ points and pad its discrete Fourier transform
with $N$ zeros,
by inverse transform we obtain a $2N$-point time series.
The second step consists in applying splines
to interpolate the upsampled time series to the barycentric
time for number of points equal to the number of original data points. Thus if the
original
time series contains $N$ points the final interpolated time series contains also
$N$ points.

\begin{figure}
\scalebox{0.5}{\includegraphics{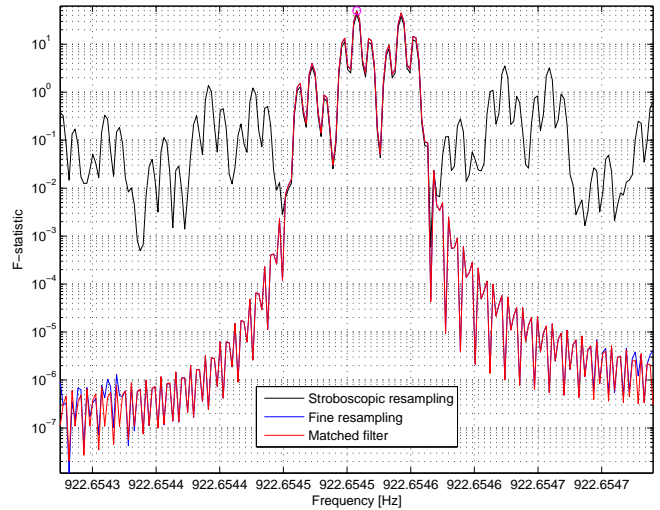}}
\caption{\label{fig:resampling} Comparison of the two interpolation methods and
the
perfect matched-filtering. We see that the two-step interpolation method that
uses Fourier
and spline interpolation very accurately reproduces the perfect matched filter.}
\end{figure}

We have compared the performance of the two interpolation methods
and we have also compared these methods with an exact matched filter.
To carry out the comparison we have used noise-free waveforms
given in Eqs.\ (\ref{resp})--(\ref{pha3}) with one spin-down parameter.
We have calculated the $\F$-statistic using the two interpolation methods
and exact matched-filtering method. In the matched filtering method we have
assumed
that we know the frequency of the signal and thus the Doppler modulation due to
the motion of the detector.
However, we have used FFT to calculate the $\F$-statistic for the whole
frequency band.
The results are shown in Fig.\ \ref{fig:resampling}.

We have performed a Monte Carlo simulation consisting of 1000 trials
and we have found that the rms error divided by maximum of the $\F$-statistic in
the second method was $0.1\%$
whereas in the first, fastest method it was $5\%$. The nearest neighbor
interpolation leads to a greater
signal-to-noise ratio loss than spline interpolation and also, very importantly,
to
elevated sidelobes of the $\F$-statistic. In the presence of noise this can lead
to a loss
of the parameter estimation accuracy if the noise elevates the sidelobes above
the main
maximum. The stroboscopic resampling is much faster than the second two step
method
however the second method is much more accurate than the first.

\subsection{FFT interpolation}
\label{sSec:Krob}

Using the FFT algorithm we can efficiently calculate
the \emph{discrete Fourier transform} (DFT) $X(k)$ ($k=1,\ldots,N$)
of a time series $x_\ell$ ($\ell=1,\ldots,N$).
We recall that $X(k)$ is given by the following expression
\be
\label{eq:DFT}
X(k) = \sum^N_{\ell = 1} x_\ell e^{-2\pi\mi(\ell-1)(k-1)/N},
\quad k=1,\ldots,N.
\ee
The frequencies $(k-1)/N$ are called {\em Fourier frequencies}
and DFT components calculated at Fourier frequencies are called {\em Fourier bins}.
When the true frequency of a monochromatic signal does not coincide with one of
the Fourier frequencies,
the use of the FFT algorithm to evaluate the sum (\ref{eq:DFT}) leads a certain
loss of signal-to-noise ratio.
The greatest loss equal around 36.3\% is when the true frequency is half way
between the Fourier frequencies.

One way to improve this situation is to pad the time series of $N$ data points
with $N$ zeros.
This leads to DFT evaluated at twice as many points as the DFT of the original
time series
and the signal-to-noise loss is only 9.97\%. However this procedure leads
to evaluating twice as long FFT as the original ones
and thus increases the computational time by more than a factor of two.

There exists an approximate interpolation procedure proposed by pulsar
astronomers
(see Chapter 7.3.3 in Ref.\ \cite{JKbook}),
in which the DFT component in the middle of two Fourier frequencies is
approximated by
\be
\label{eq:intbin}
X(k + 1/2) \cong [X(k+1) - X(k)]/\sqrt{2}.
\ee
This interpolation method is called {\em interbinning}.
One can show (see Fig.\ 7.3 in Ref.\ \cite{JKbook})
that the interpolation based on Eq.\ \eqref{eq:intbin}
leads to maximum loss of signal-to-noise ratio of 13\%.

\subsection{Finding the maximum of the $\F$-statistic accurately}
\label{sSec:Kroc}

\begin{figure}
\scalebox{0.5}{\includegraphics{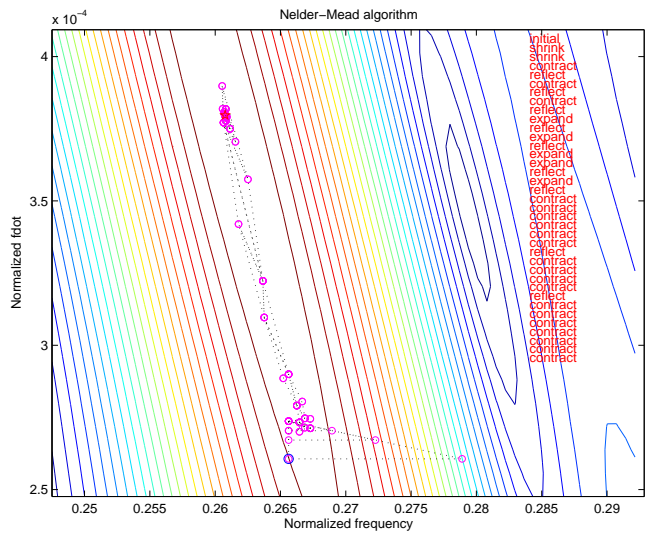}}
\caption{\label{fig:nm}
An illustration of the Nelder-Mead algorithm in two dimensions.
We first evaluate the function to be minimized on an initial triangle.
At each step of the search, a new point in or near the current triangle
is generated. The function value at the new point is compared with
the function's values at the vertices of the triangle and, usually, one of
the vertices is replaced by the new point, giving a new triangle. This step
is repeated until the diameter of the simplex is less than the specified
tolerance.
In the method the triangle adapts itself to the local landscape, elongating down
long inclined planes,
changing direction on encountering a valley at an angle, and contracting in the
neighborhood of a minimum.}
\end{figure}

As we calculate the $\F$-statistic on a discrete grid in the parameter space
and as the maximum of the $\F$-statistic does not in general coincide with a
node of the grid,
we always lose some signal-to-noise ratio and parameter estimation accuracy.
To improve this we use a nonlinear optimization routine
to find an improved maximum of our statistic.

The search for the maximum of the $\F$-statistic is performed in two steps.
First we find the maximum of $\F$ over the discrete grid in the parameter space
and then the parameters obtained in this {\em coarse search} we input as initial values
to some hill-climbing optimization routine to find an improved maximum.
The second step is called a {\em fine search}. As our maximum finding routine
we can use a direct search that does not require calculations of derivatives of
$\F$,
namely the Nelder-Mead algorithm or simplex search algorithm \cite{nelder-65}.
The Nelder-Mead algorithm is illustrated in Fig.\ \ref{fig:nm}
for two-dimensional case where simplices are triangles.

\section{Monte Carlo simulations}
\label{Sec:Kro3}

We have implemented the data analysis methods presented in the previous sections
in a computer code and we have performed a number of Monte Carlo simulations
to test how accurately we can estimate signal's parameters
for different values of the signal-to-noise ratio (SNR)
and how the rms errors of our ML estimators
compare to the Cram\`er-Rao lower bound. For unbiased estimators
the Cram\`er-Rao lower bound on variances of the estimators is given by the diagonal elements of
the inverse of the Fisher matrix. The ML estimators are asymptotically
(i.e.\ when SNR tends to infinity) unbiased and with variances approaching the
diagonal elements of the inverse of the Fisher matrix.
Our Monte Carlo simulations consisted of generating the data that were the sum
of the white noise and the gravitational-wave signal presented in Sec.\ \ref{Sec:Pio}
and using the filtering procedure from Sec.\ \ref{Sec:Kro1}
to detect the signal and estimate its parameters.
We have taken the observation time equal to exactly 2 sidereal days.
For this observation time it is enough to take only the first spin-down
parameter in the templates (see Refs.\ \cite{BCCS98,JK00}).
To obtain the Cram\`er-Rao lower bound we have calculated the $8 \times 8$ Fisher matrix
[defined in Eq.\ \eqref{Fm-def}] for the signal model given in Sec.\ \ref{Sec:Pio}
with one spin down parameter included.

We have made our simulation to mimic the analysis of the VIRGO interferometer VSR1 data.
We have thus assumed that data comes from an interferometric detector
located at the position of the VIRGO detector
and we have used the detector's ephemerides corresponding to data taken sometime in the year 2007.
We have generated data in a narrow frequency band of 50~mHz
with the lower edge of the band equal to 435.1875~Hz.
This resulted in a short time series of 17233 data points
greatly reducing the CPU time needed to perform the simulations.
For all the simulations we have chosen
the same parameters of the gravitational-wave signal
except for the constant amplitude $h_0$ that we scaled to obtain data
with a chosen SNR [defined in Eq.\ \eqref{snr-def}].
We have found the grid point $p_g$ nearest to the true position of the signal in
the parameter space and we have calculated the $\F$-statistic on a small grid around the point $p_g$.
The size of the grid was $\pm2$ grid points from the point $p_g$ in the direction
of the parameters $\dot{\omega}$, $\alpha_1$, $\alpha_2$ [see Eqs.\ \eqref{eq:albe}].
For each set of these three parameters we have evaluated the $\F$-statistic
for the whole 50~mHz band.

In all the simulations we have used
the constrained grid constructed in Sec.\ \ref{Sec:Mac}
and we have chosen the detection threshold for the $\F$-statistic equal to 10.
This low threshold ensured that the probability of detection
was nearly one and none of the signals was missed.
In each simulation the estimation of signal's parameters was performed in two steps.
The first step, called the {\em coarse} search,
was the calculation of the $\F$-statistic on the grid described in Sec.\ \ref{Sec:Mac}.
We have registered all the threshold crossings
and we have taken the coarse estimates of the signal's parameters
as the parameters of the grid point for which the $\F$-statistic was maximal.
In the second step, called the {\em fine} search,
we have found the maximum of the $\F$-statistic for each signal registered in the first step
using the Nelder-Mead algorithm with the initial values equal to the coarse estimates of the parameters.
In this step the $\F$-statistic was calculated [by means of Eqs.\ \eqref{OS}--\eqref{Fab}]
exactly without any approximations.
For each value of the SNR we have repeated the simulation 1000 times
with different realizations of the white Gaussian noise.

\begin{figure*}
\begin{tabular}{lr}
\includegraphics[scale=0.40]{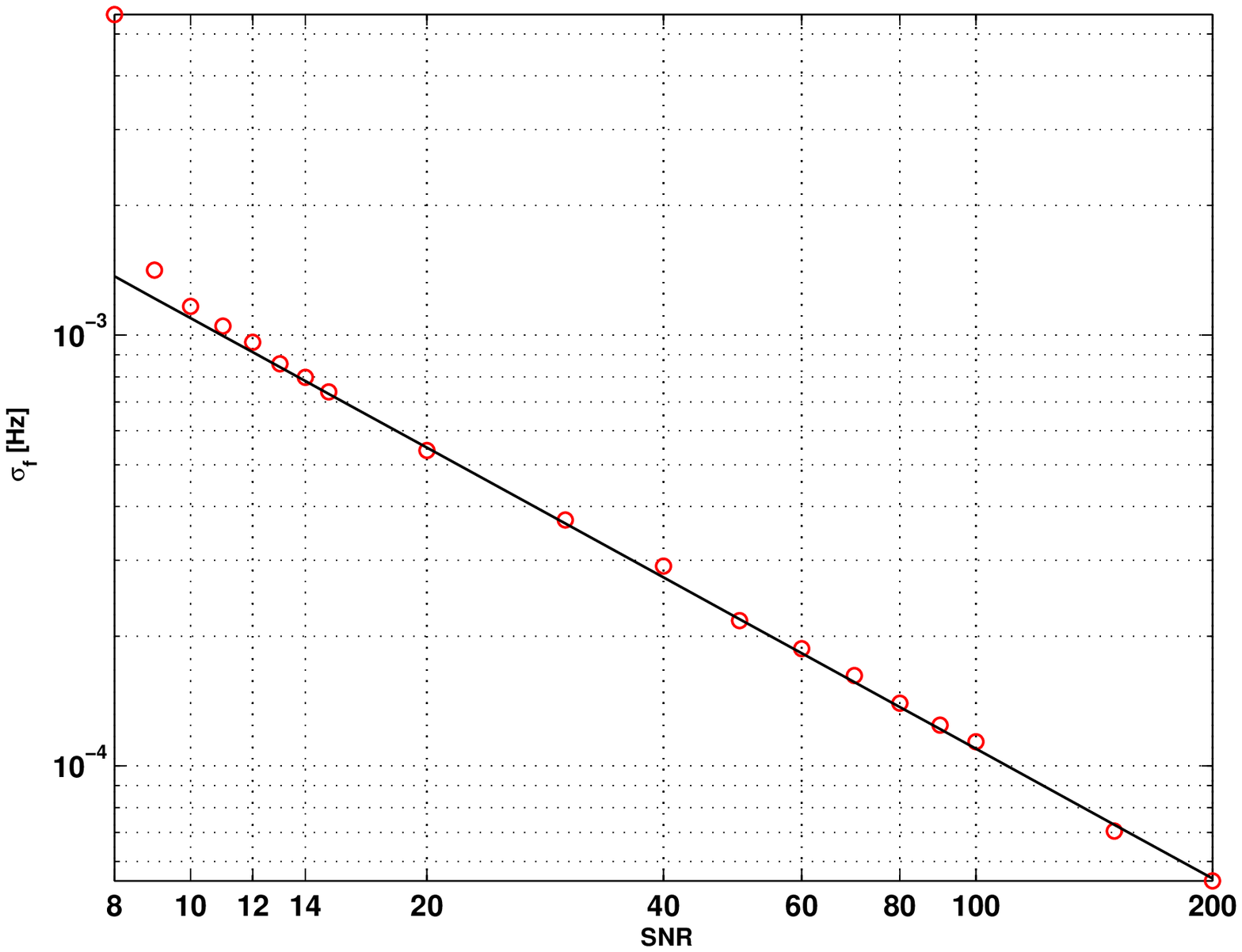} &
\includegraphics[scale=0.40]{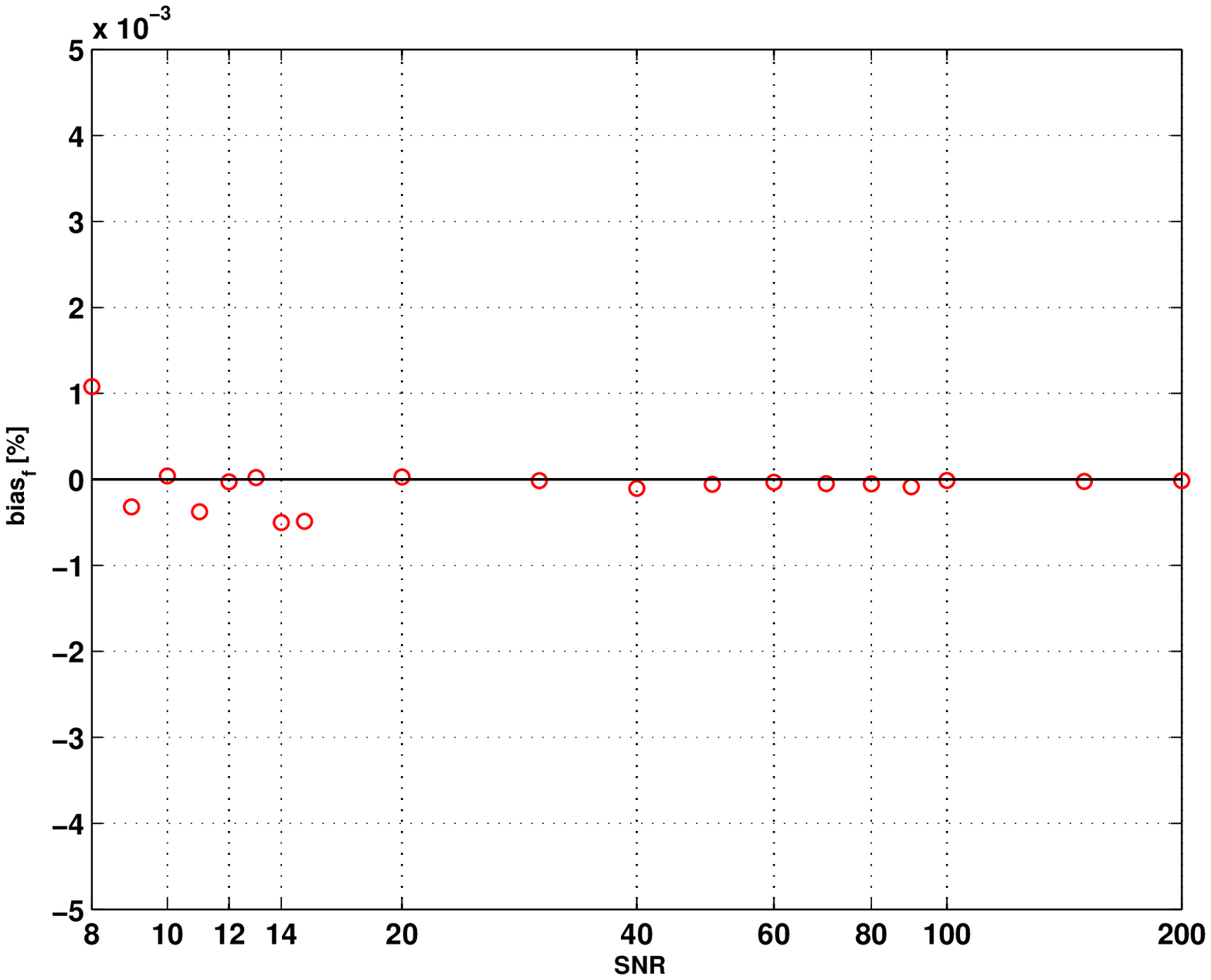}\\[1ex]
\includegraphics[scale=0.40]{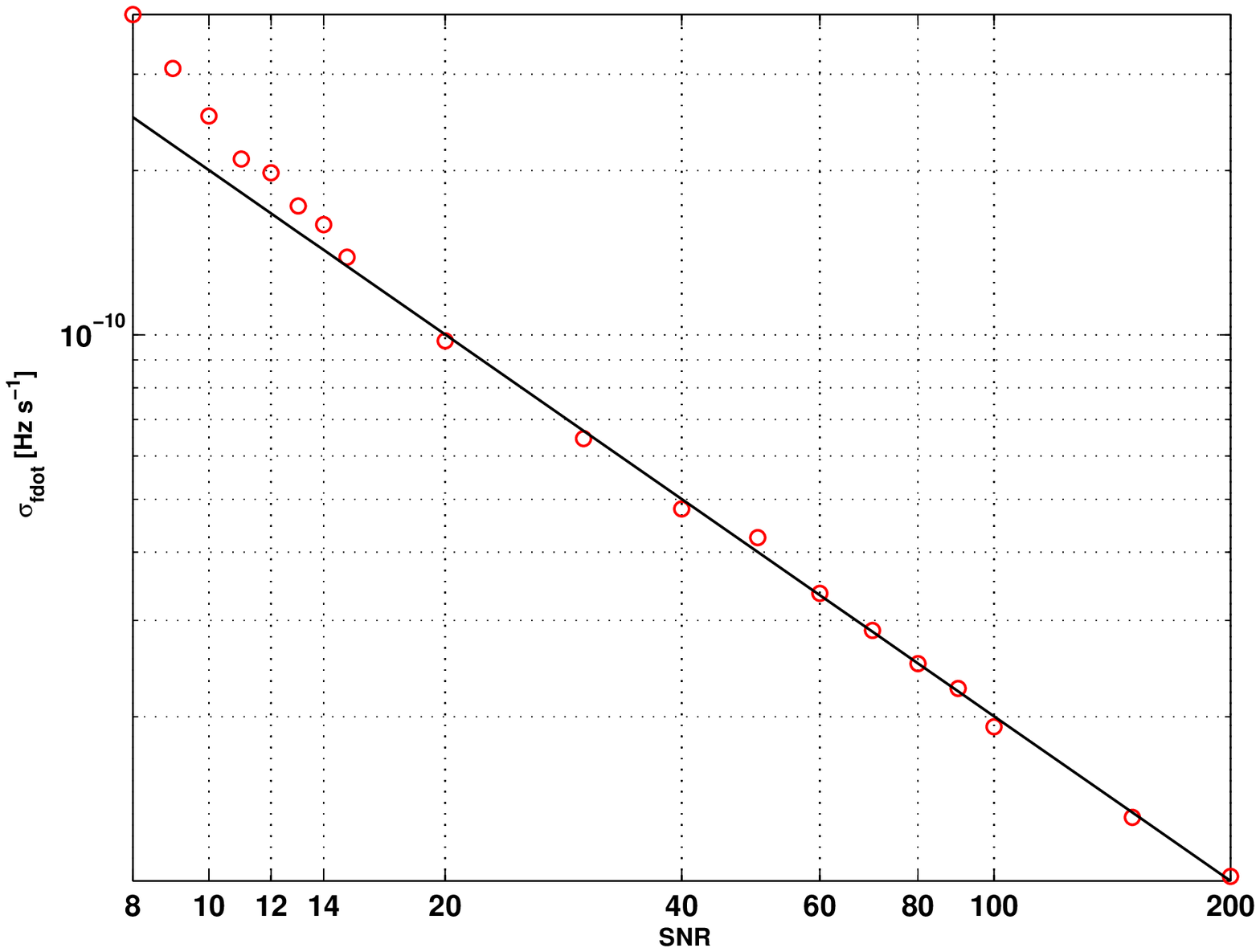} &
\includegraphics[scale=0.40]{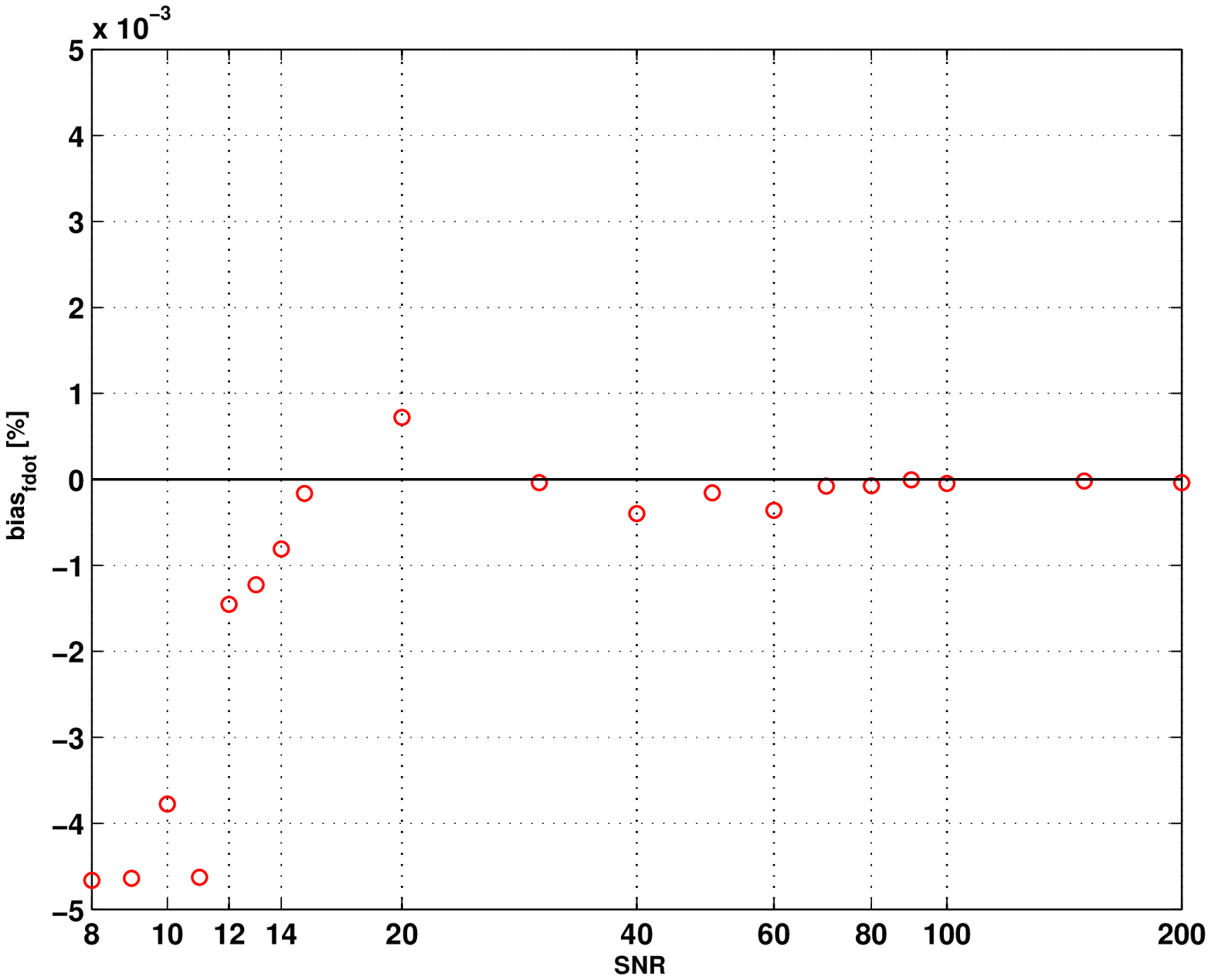}\\[1ex]
\includegraphics[scale=0.40]{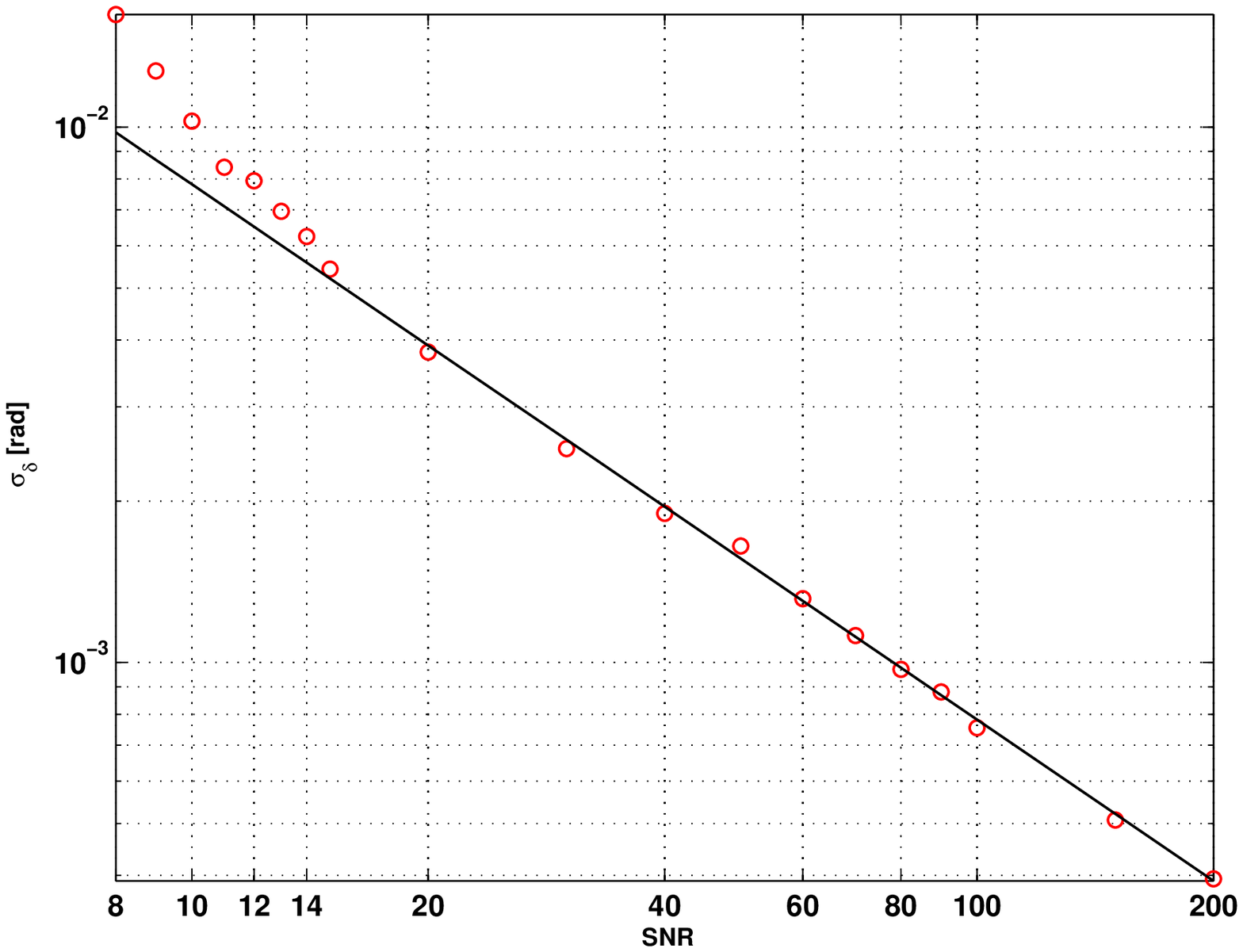} &
\includegraphics[scale=0.40]{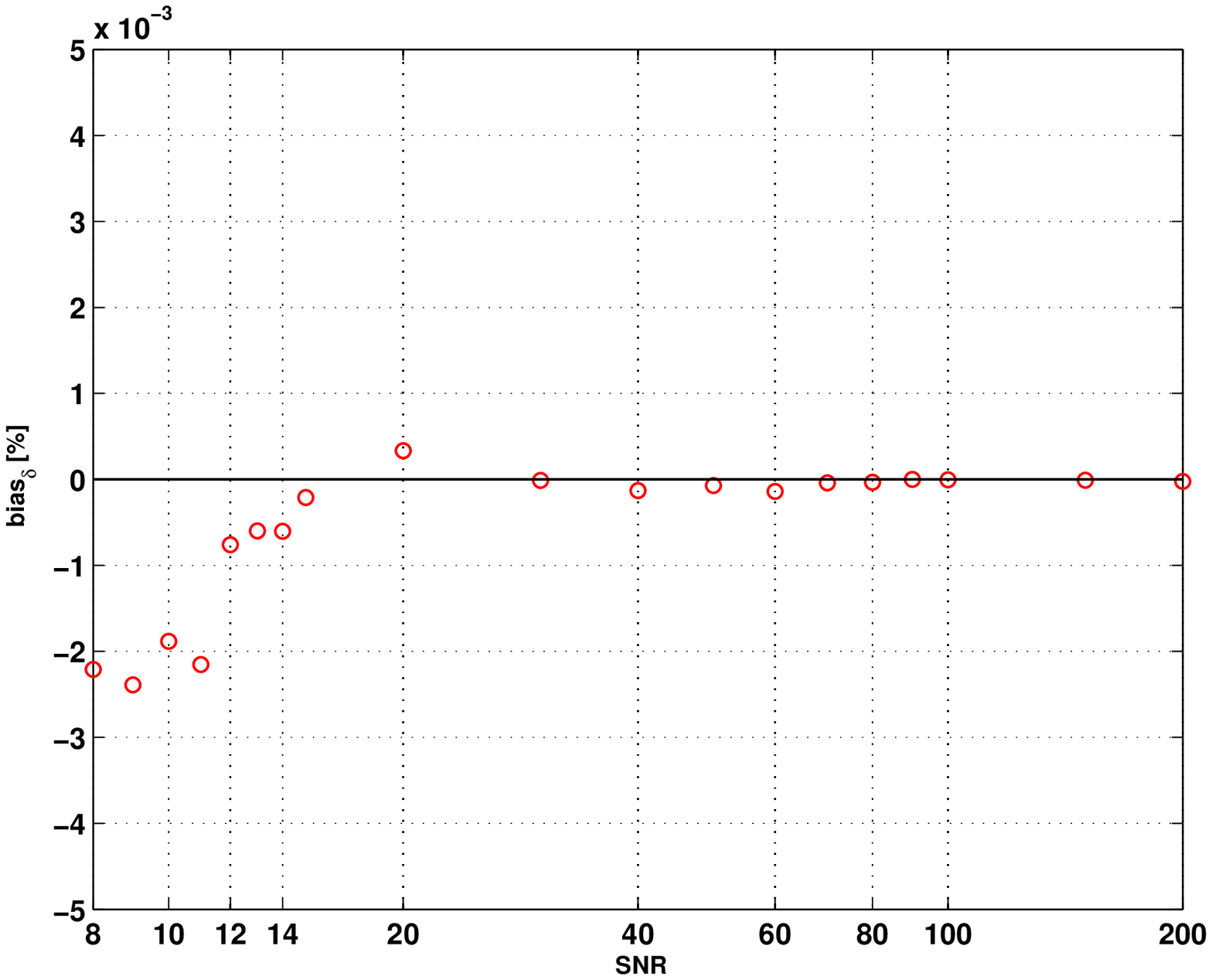}\\[1ex]
\includegraphics[scale=0.40]{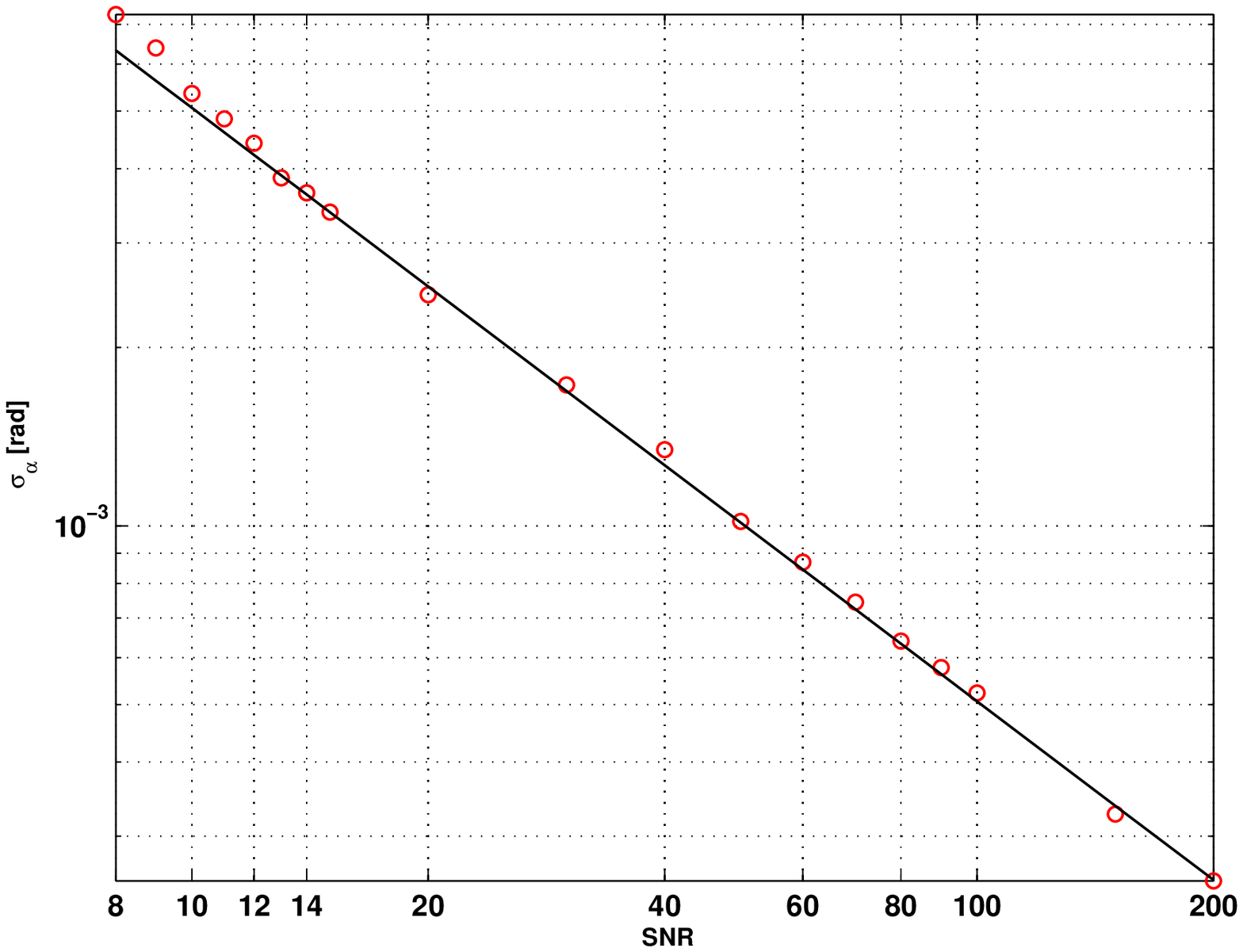} &
\includegraphics[scale=0.40]{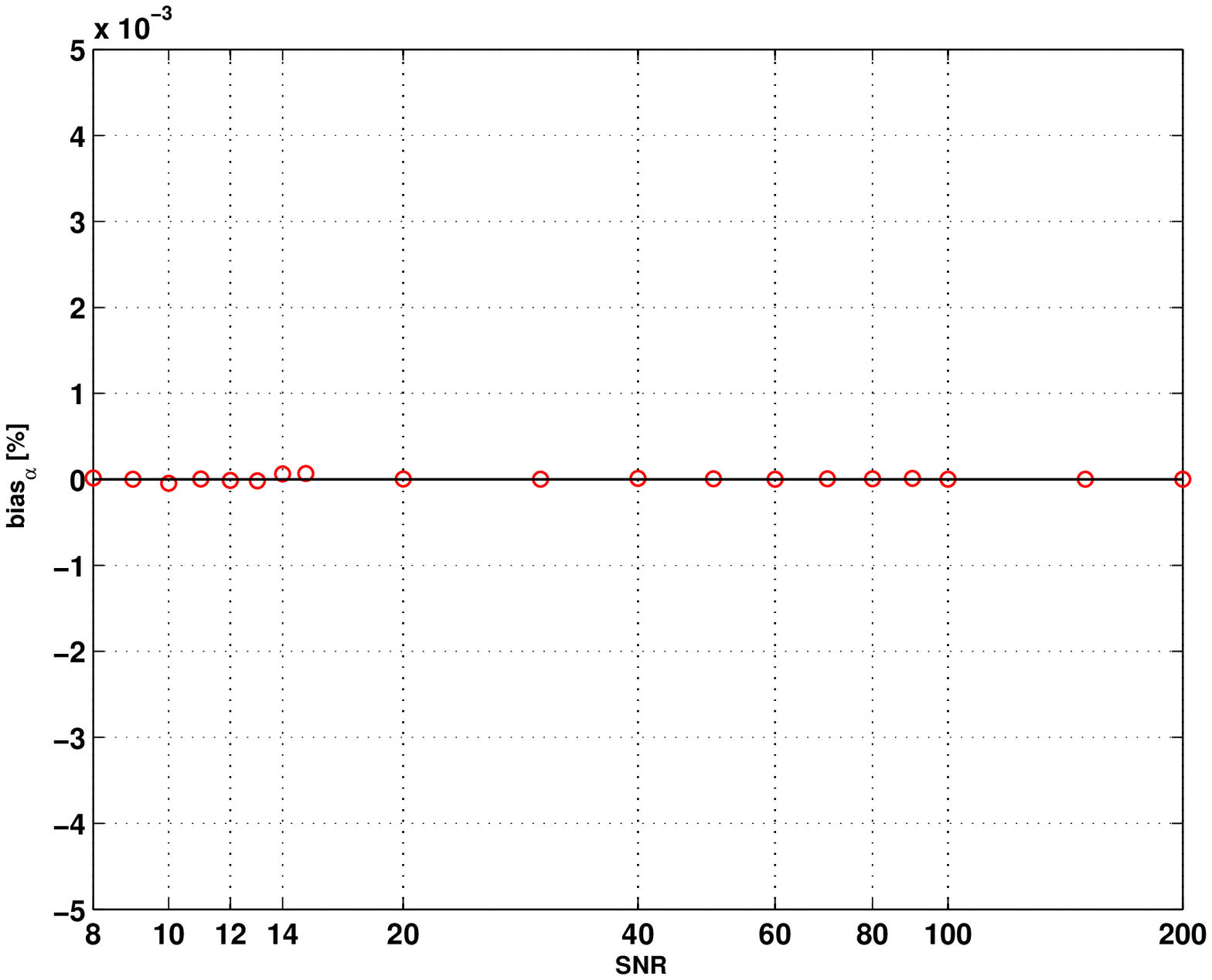}
\end{tabular}
\caption{\label{fig:sig}
Standard deviations (left panels) and biases (right panels) of the ML estimators
of the intrinsic parameters as functions of the SNR
computed with the grid of $\mathrm{MM}=\sqrt[3]{0.9}$.
The plots are made, from the top, for the following parameters:
frequency, spin down, declination, and right ascension.
Each circle corresponds to a standard deviation
or a bias computed from simulation of 1000 runs.
The continuous lines in the left panels are the Cram\'er-Rao bounds
for standard deviations.}
\end{figure*}

In the first simulation we have employed a fine grid
with the minimal match $\mathrm{MM}=\sqrt[3]{0.9}$.
We have used the accurate two-step resampling procedure
described in Sec.\ \ref{sSec:Kroa} and before applying the FFT
we have padded the data with 15535 zeros resulting in a time series of $2^{15}$ points.
This ensured an exact interpolation of the DFT between the Fourier frequencies
and the fastest implementation of the FFT algorithm
(because the number of data points was a power of 2).
The results of this simulation are presented in Fig.\ \ref{fig:sig},
where we have depicted standard deviations and biases
of the \emph{intrinsic} parameters of the signal:
frequency, spin down, declination, and right ascension, as functions of the SNR.
The results of simulation of 1000 runs are marked by the circles.
Additionally in the left panels we have shown the Cram\'er-Rao bounds
for the standard deviations calculated from the inverse of the Fisher information matrix.
We see that for the SNRs greater than $\sim$9 the standard deviations obtained from simulation
are very close to the Cram\'er-Rao bounds
and the simulated biases are small fractions of a percent of the true values.

\begin{figure*}
\begin{tabular}{lr}
\includegraphics[scale=0.40]{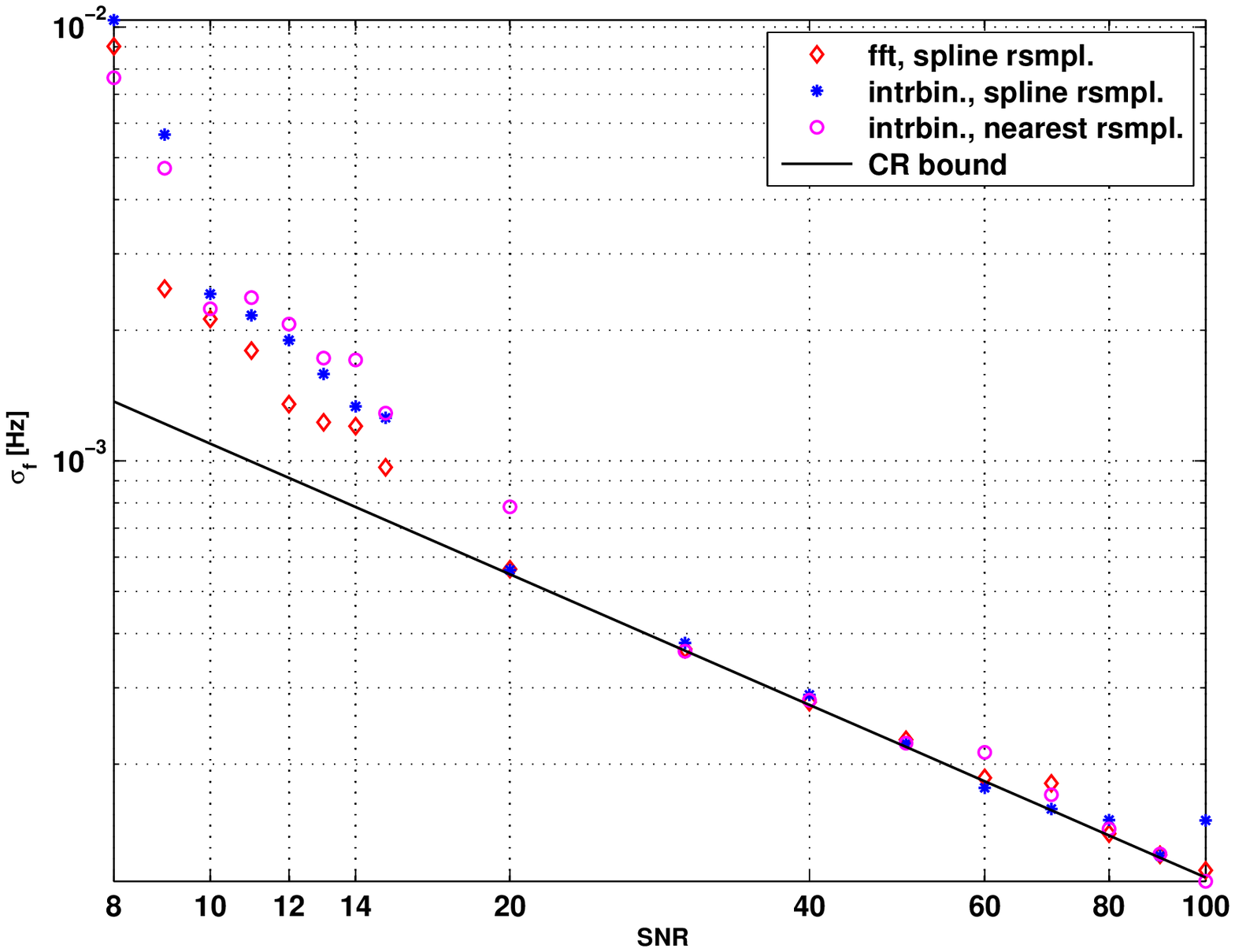} &
\includegraphics[scale=0.40]{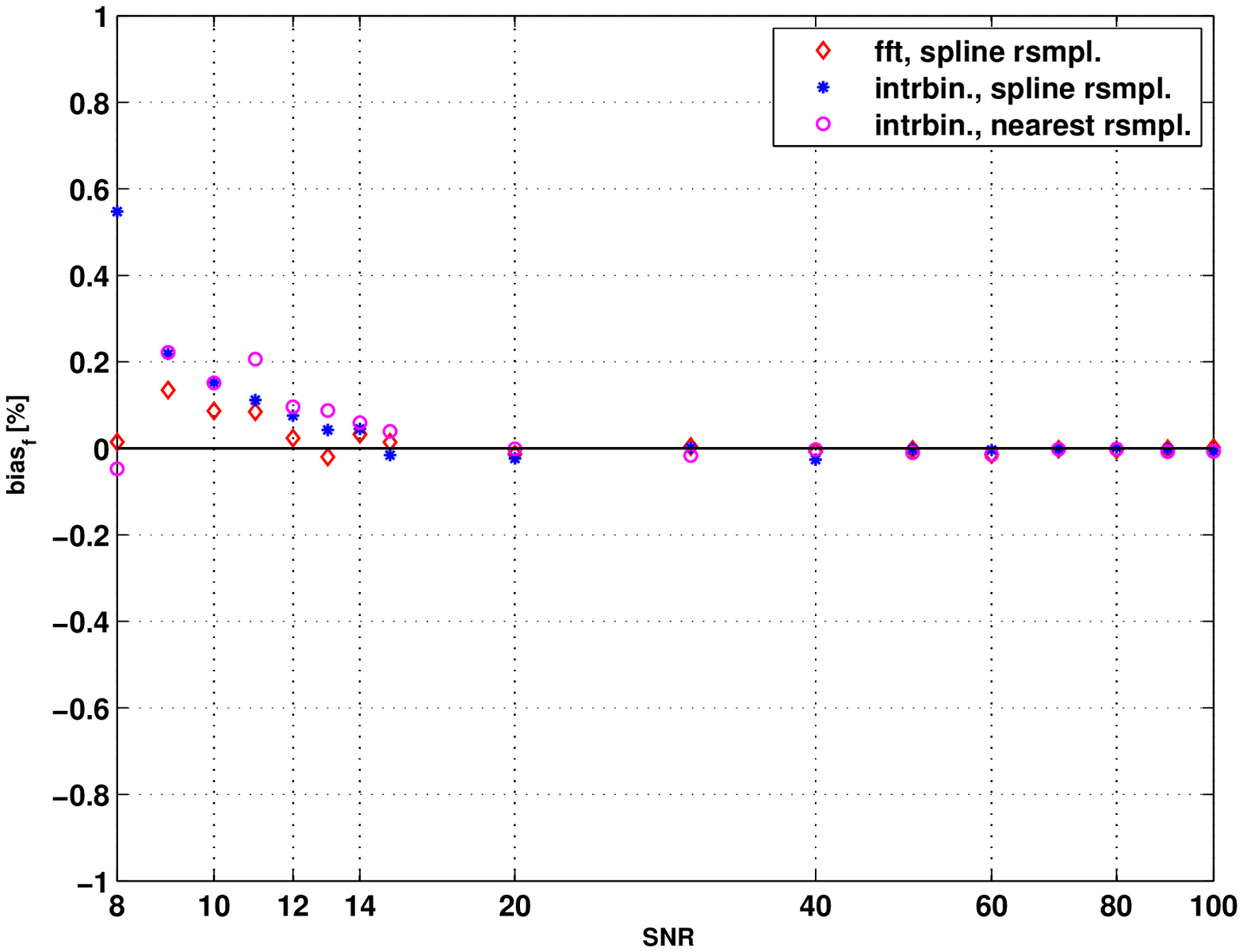}\\[1ex]
\includegraphics[scale=0.40]{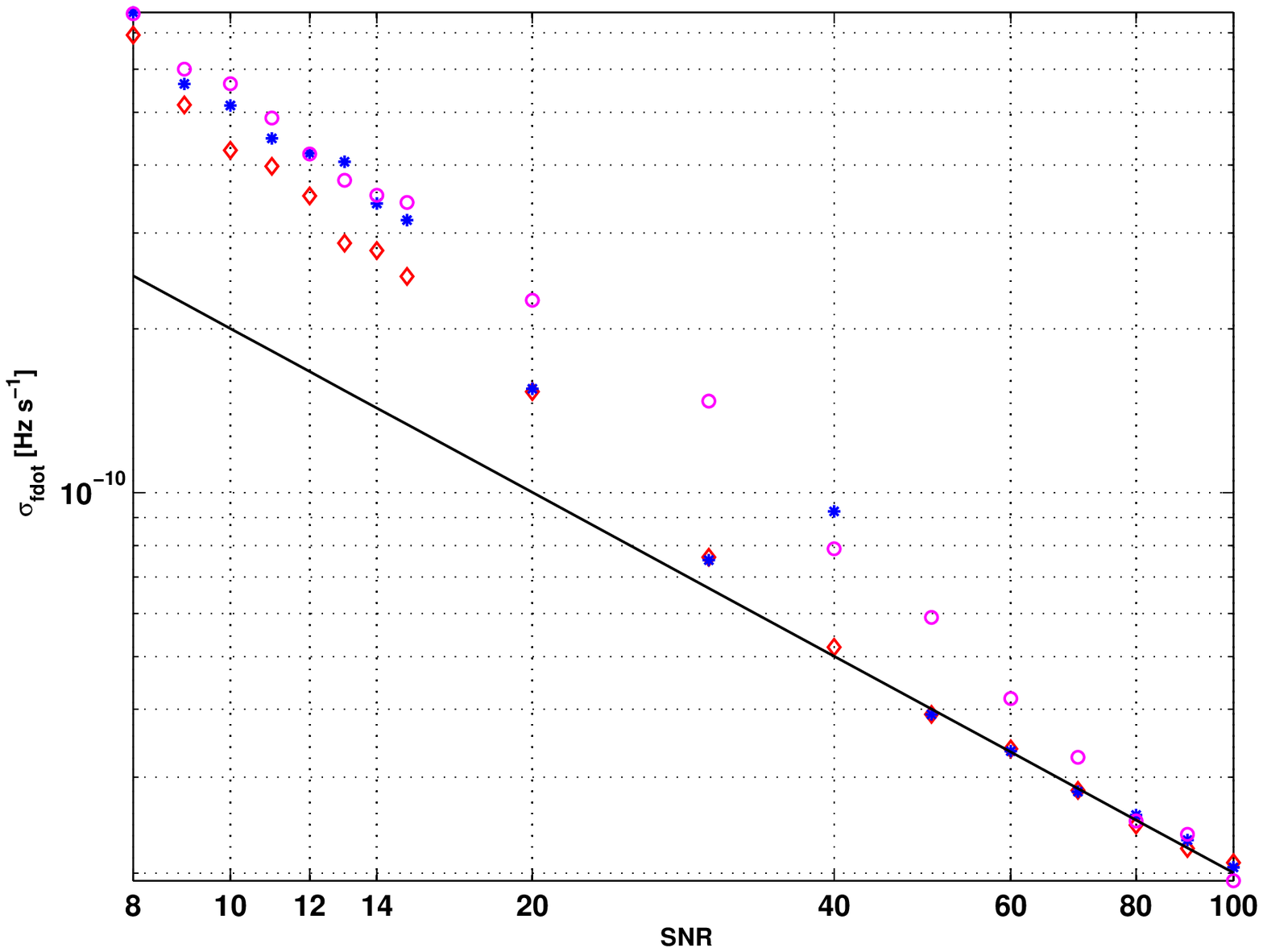} &
\includegraphics[scale=0.40]{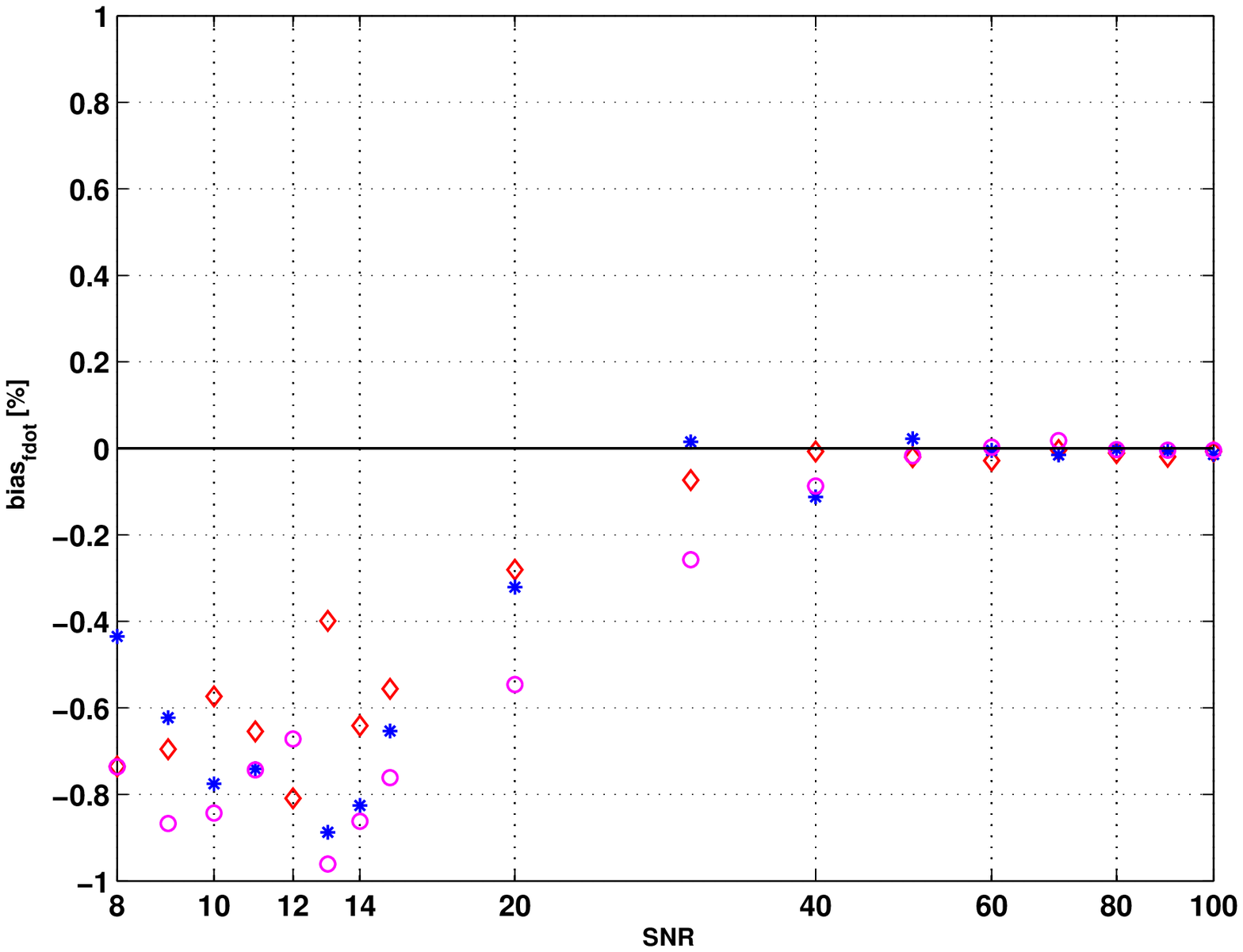}\\[1ex]
\includegraphics[scale=0.40]{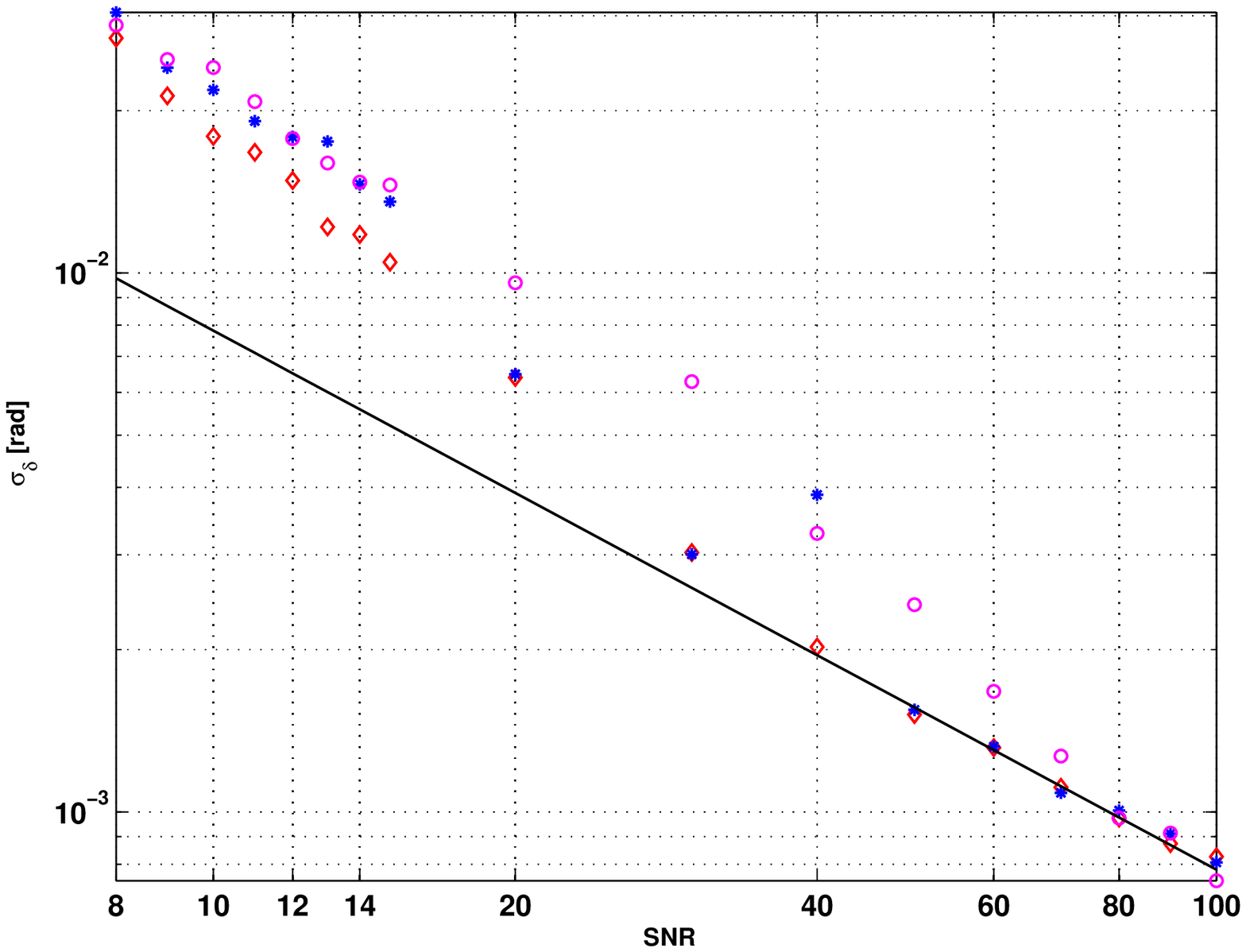} &
\includegraphics[scale=0.40]{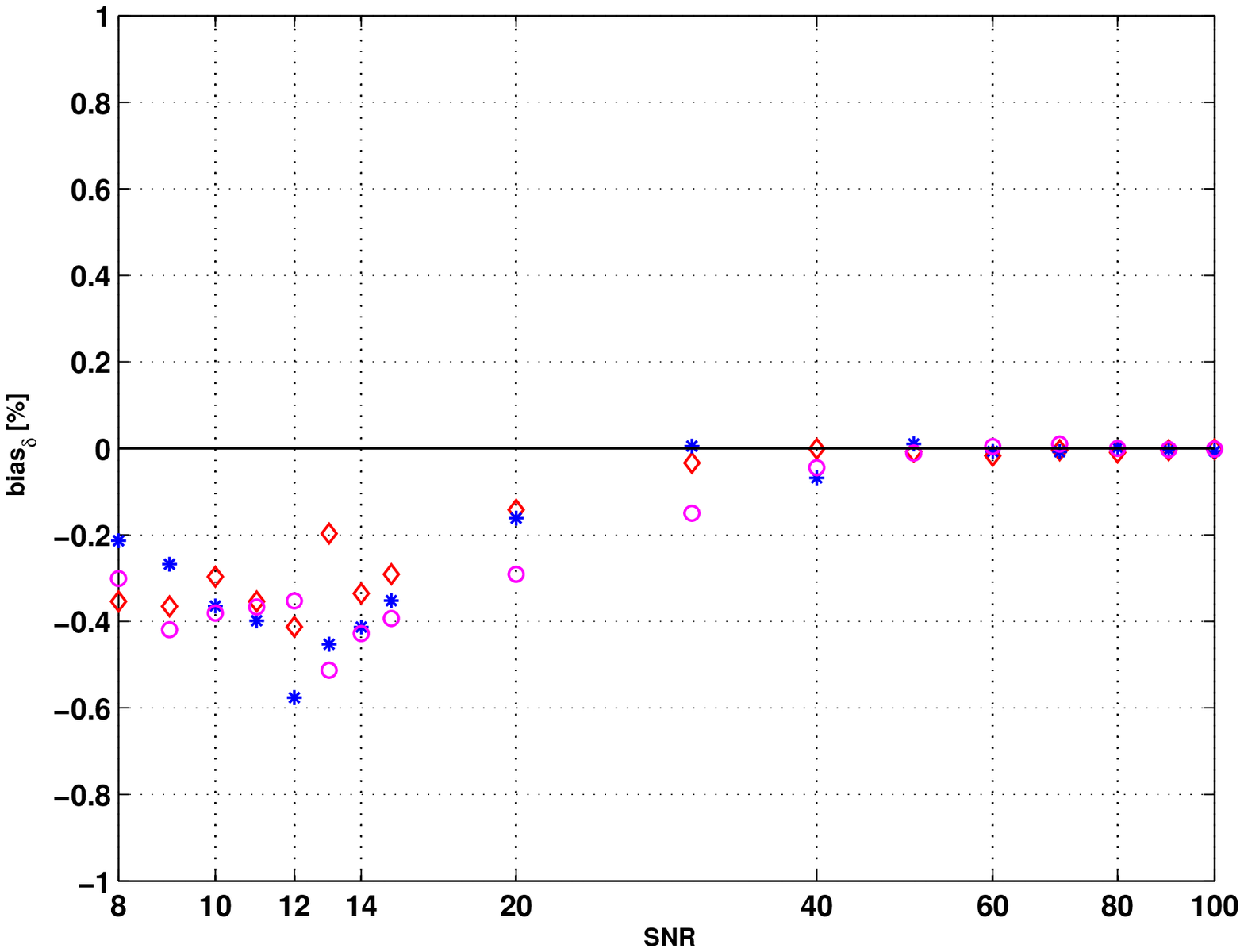}\\[1ex]
\includegraphics[scale=0.40]{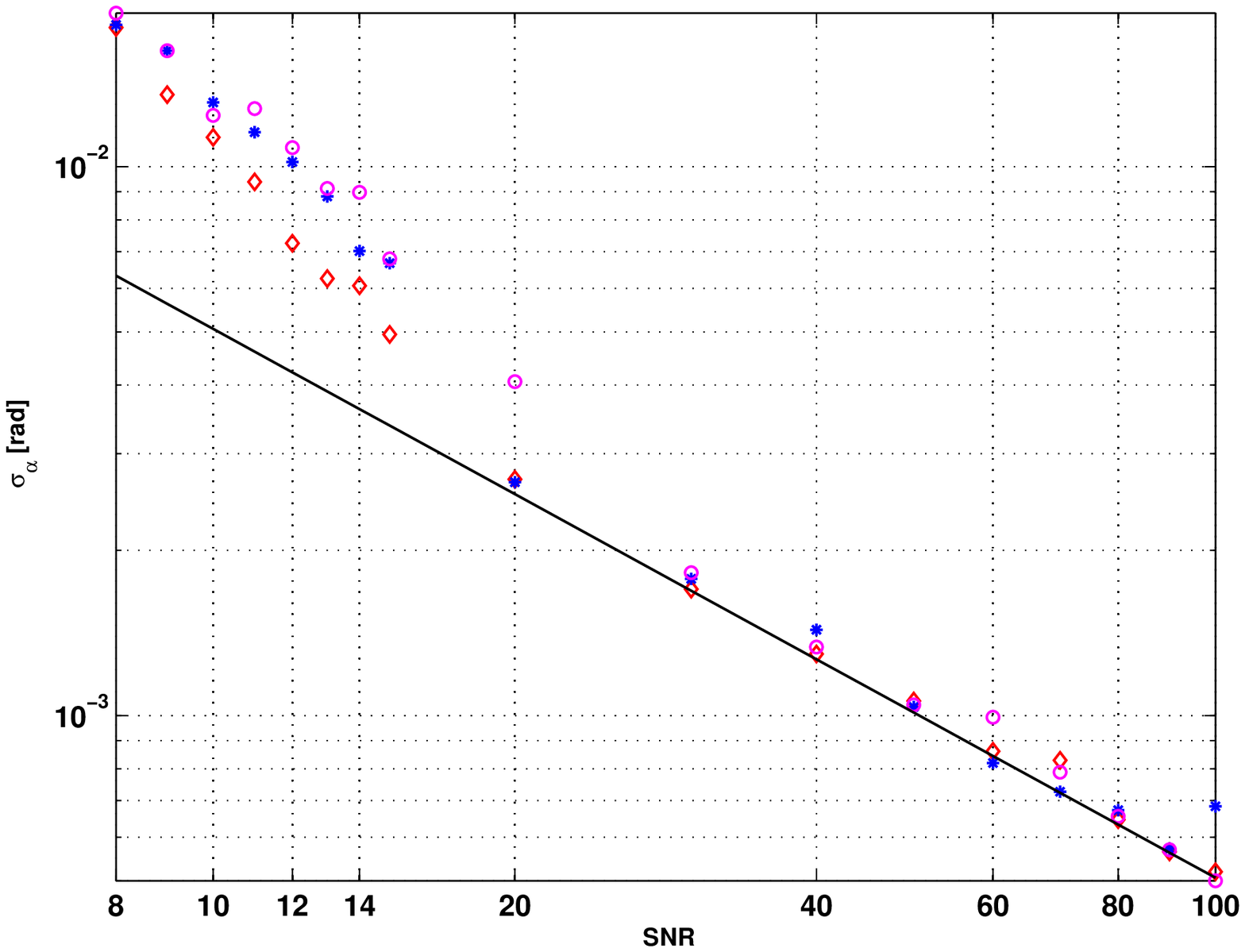} &
\includegraphics[scale=0.40]{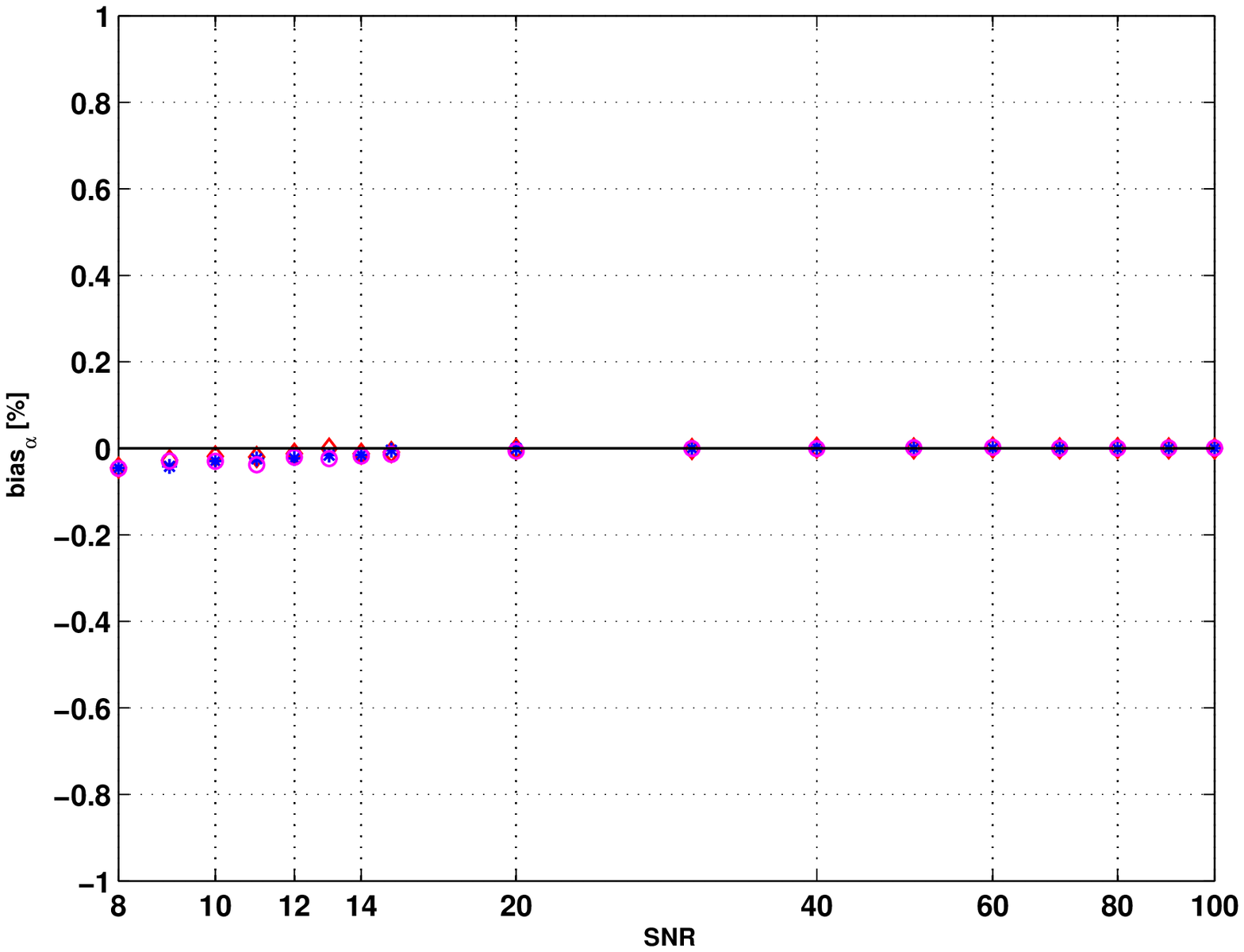}
\end{tabular}
\caption{\label{fig:sig1}
Standard deviations (left panels) and biases (right panels) of the ML estimators
of the intrinsic parameters as functions of the SNR
computed with the grid of $\mathrm{MM}=\sqrt{3}/2$ for the following three cases:
(i) zero padding and spline interpolation (diamonds);
(ii) interbinning and spline interpolation (stars);
(iii) interbinning and the nearest neighbor interpolation (circles).
The continuous lines in the left panels are the Cram\`er-Rao bounds.}
\end{figure*}

In the second simulation we have tested how accuracy of the parameter estimation
is affected by various options of the algorithms described in the previous section.
All these options aim at speeding up computations.
In the simulation we were changing the thickness of the grid,
we were comparing the two-step spline resampling with the nearest neighbor resampling,
and we were also testing the interbinning interpolation
[defined in Eq.\ \eqref{eq:intbin}].
In all the runs we have used a thicker grid corresponding
to the minimal match $\mathrm{MM}=\sqrt{3}/2$.
We have studied three specific cases:
(i) zero padding and spline interpolation;
(ii) interbinning and spline interpolation;
(iii) interbinning and the nearest neighbor interpolation.
The results of the simulation are presented in Fig.\ \ref{fig:sig1}.
{From} comparison of the three cases it follows that with a coarser grid
and with the use of different approximations
the rms errors of the parameter estimators are greater but still at a reasonable level.
In the range of the SNRs from $\sim$9 to $\sim$15,
the standard deviations obtained from simulation are twice as large as the
Cram\'er-Rao bounds, when we use spline resampling.
For the nearest neighbor interpolation, for some parameters,
the simulated rms error is twice as large as the Cram\'er-Rao bound for the SNRs up to $\sim$30.

\begin{figure*}
\begin{tabular}{lr}
\includegraphics[scale=0.40]{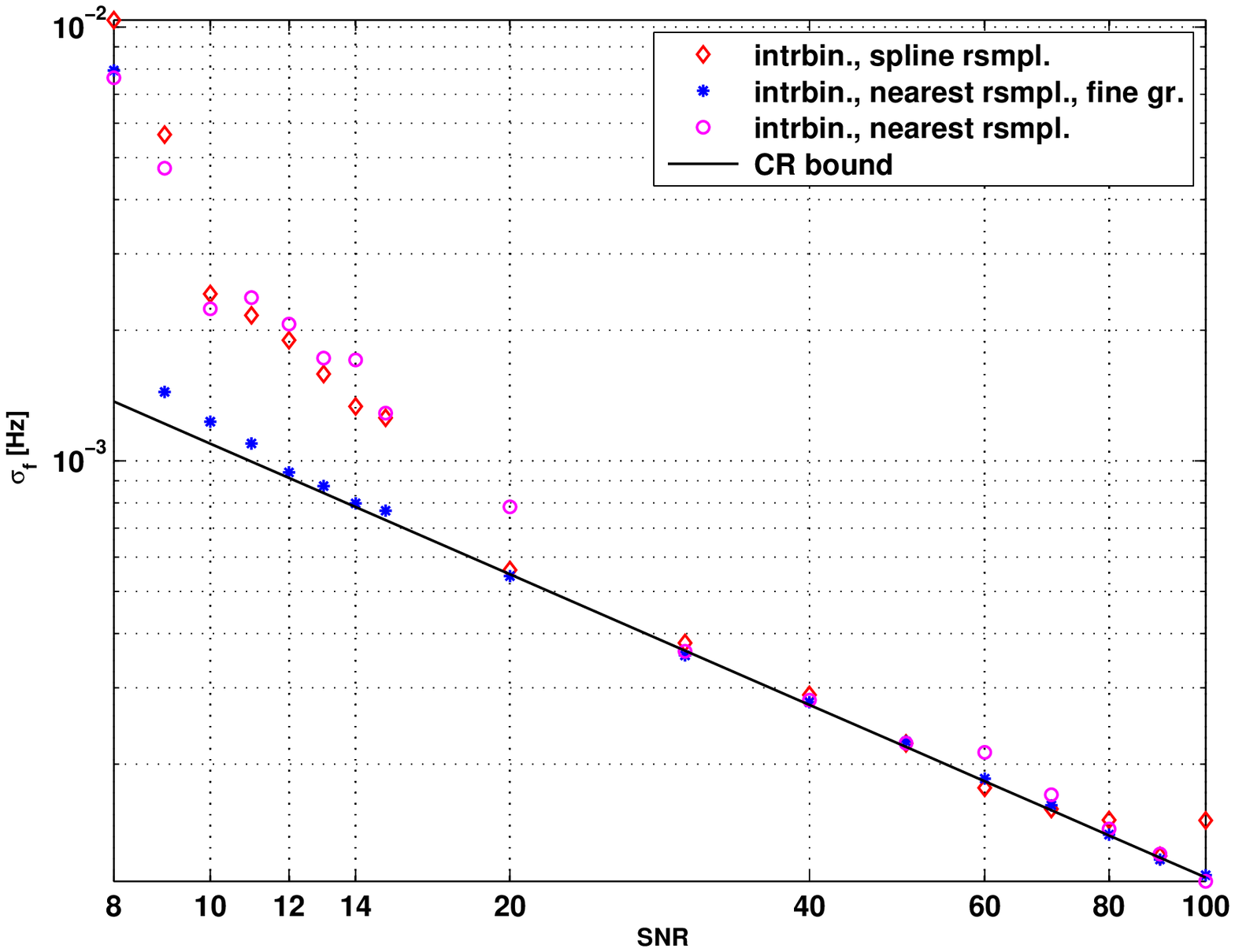} &
\includegraphics[scale=0.40]{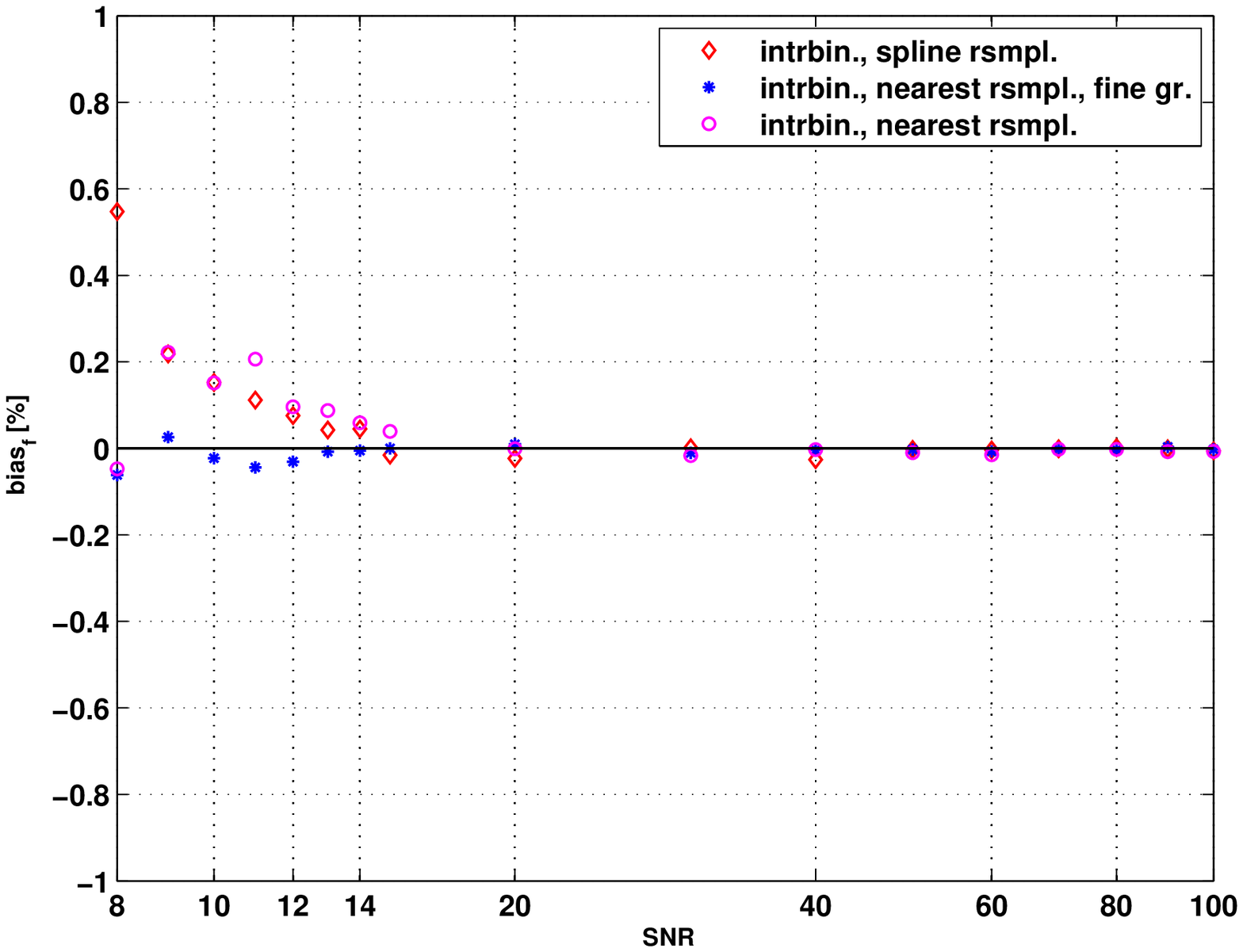}\\[1ex]
\includegraphics[scale=0.40]{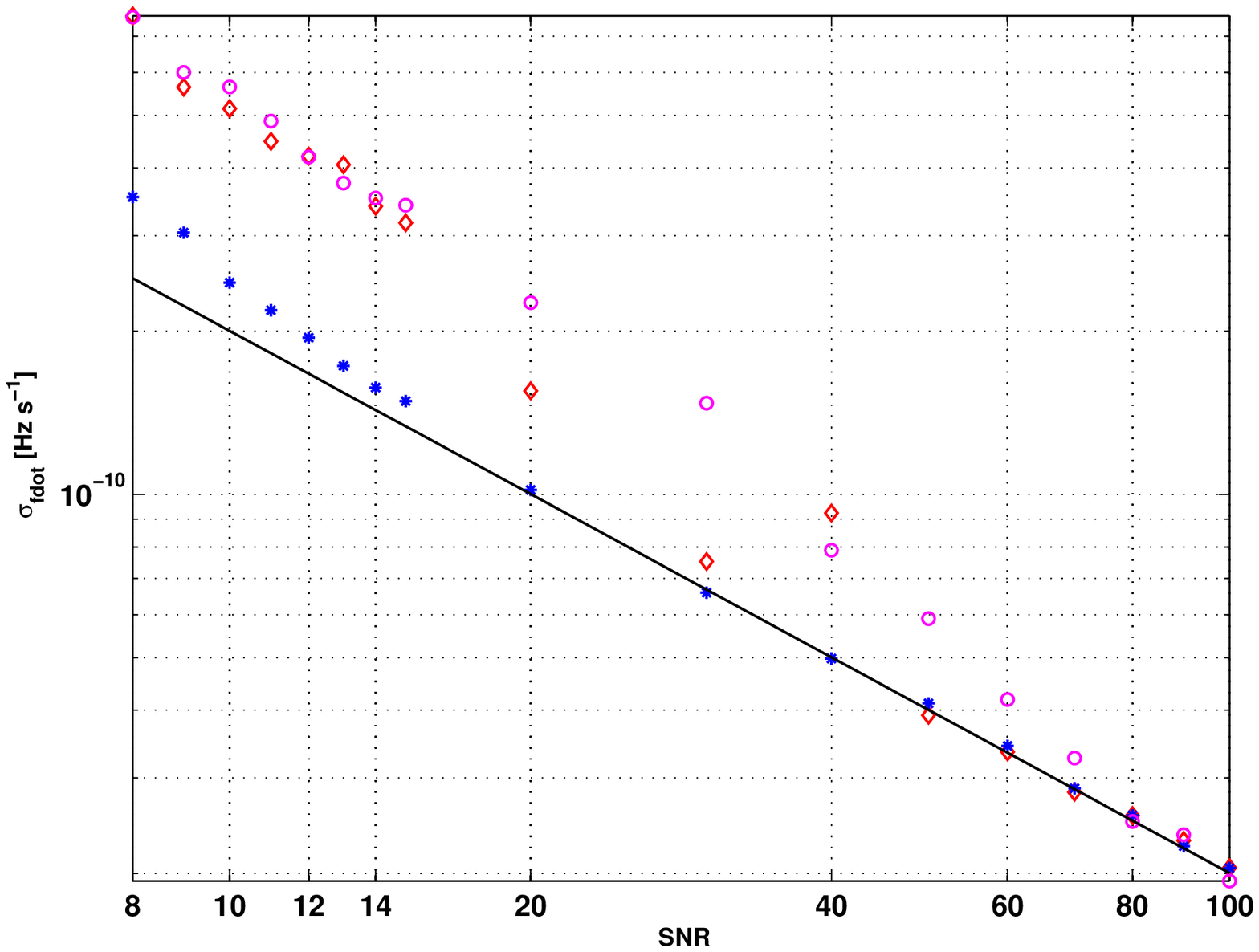} &
\includegraphics[scale=0.40]{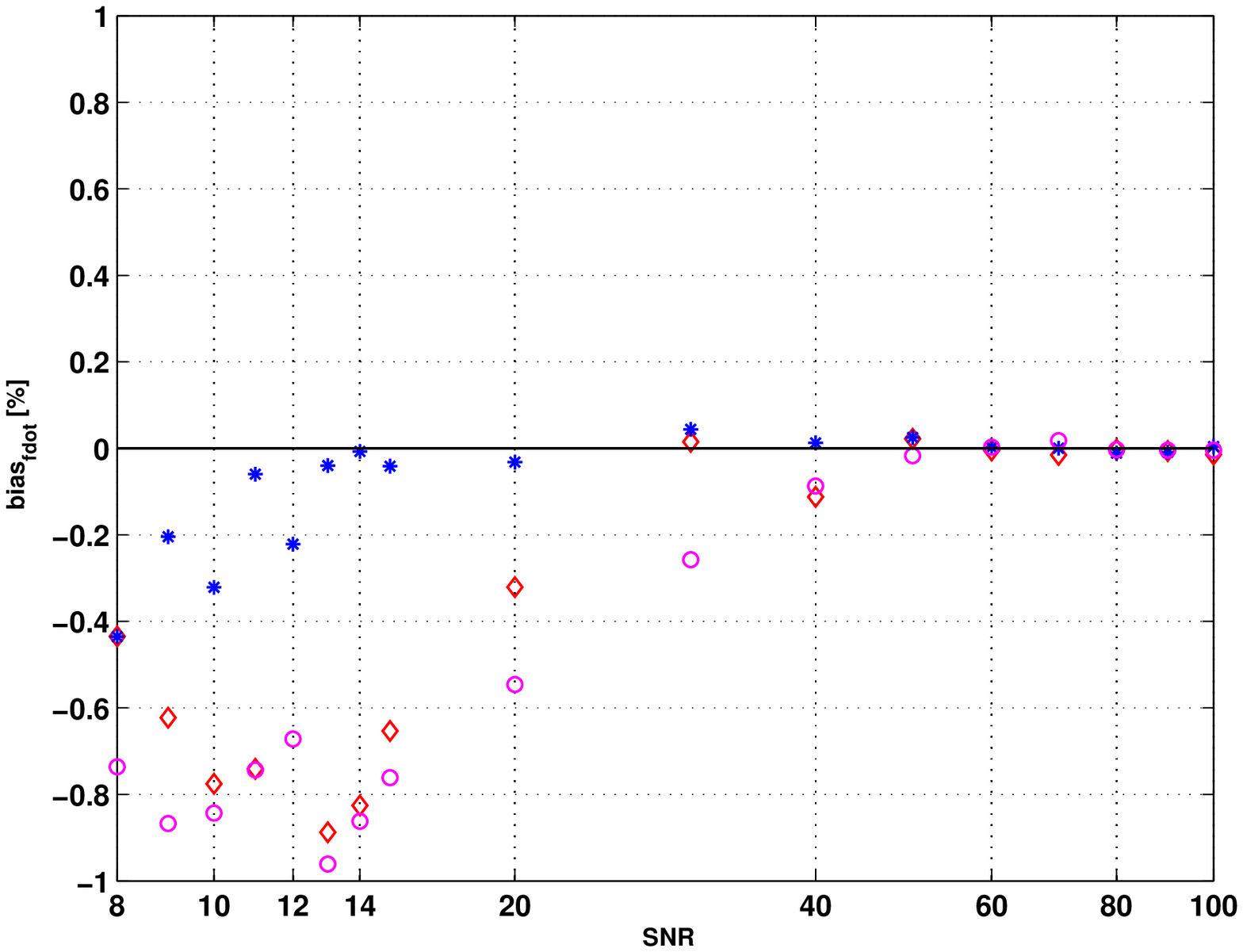}\\[1ex]
\includegraphics[scale=0.40]{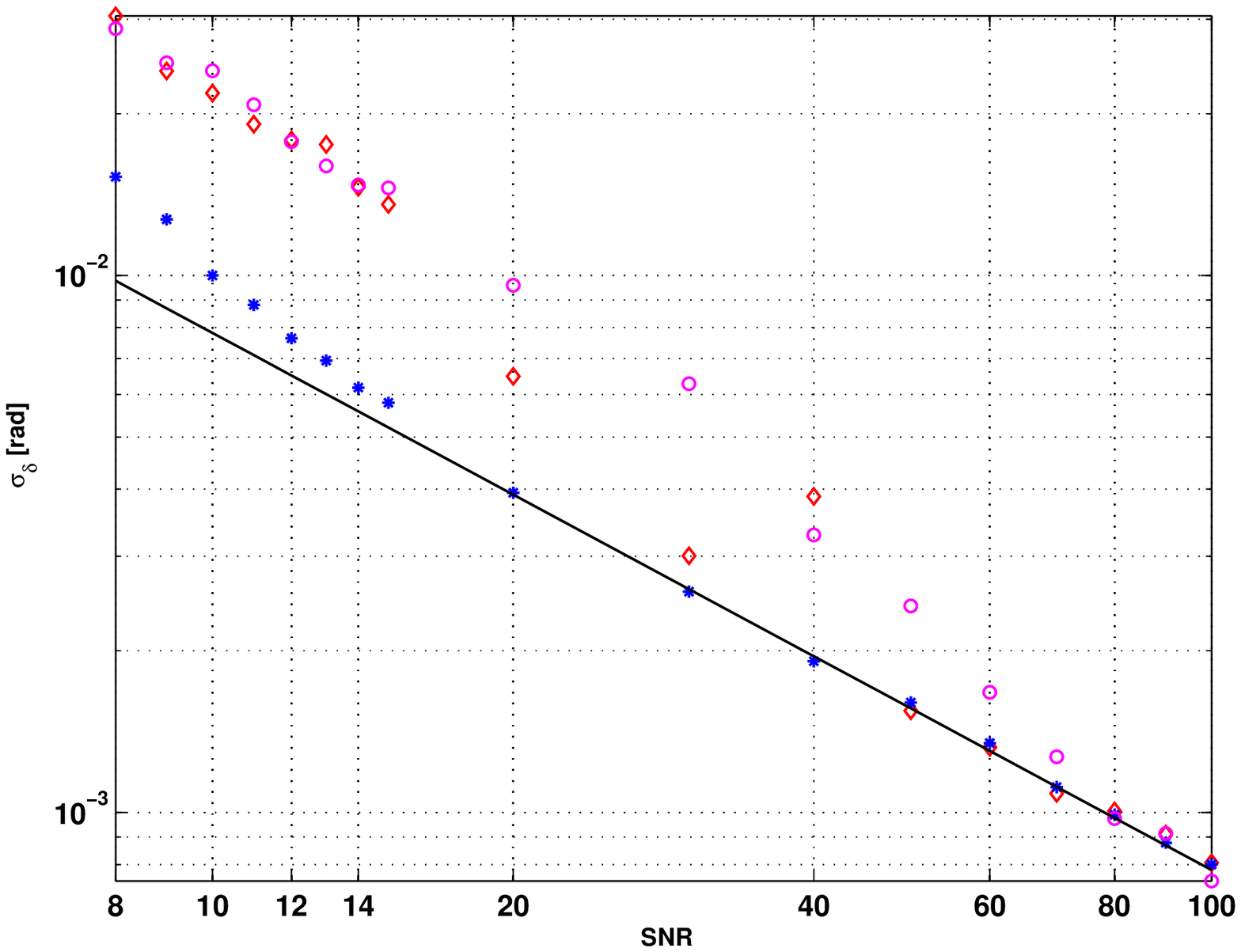} &
\includegraphics[scale=0.40]{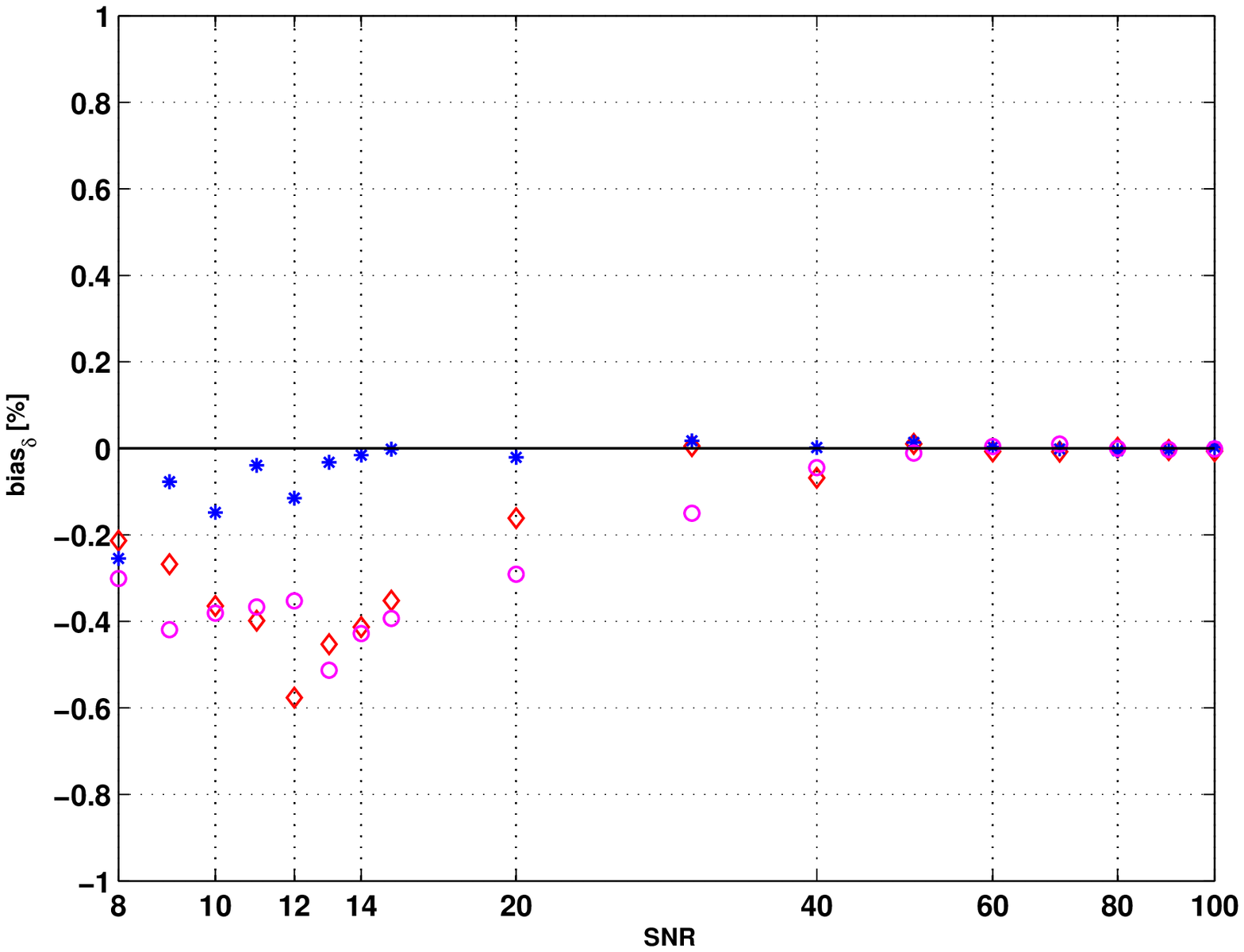}\\[1ex]
\includegraphics[scale=0.40]{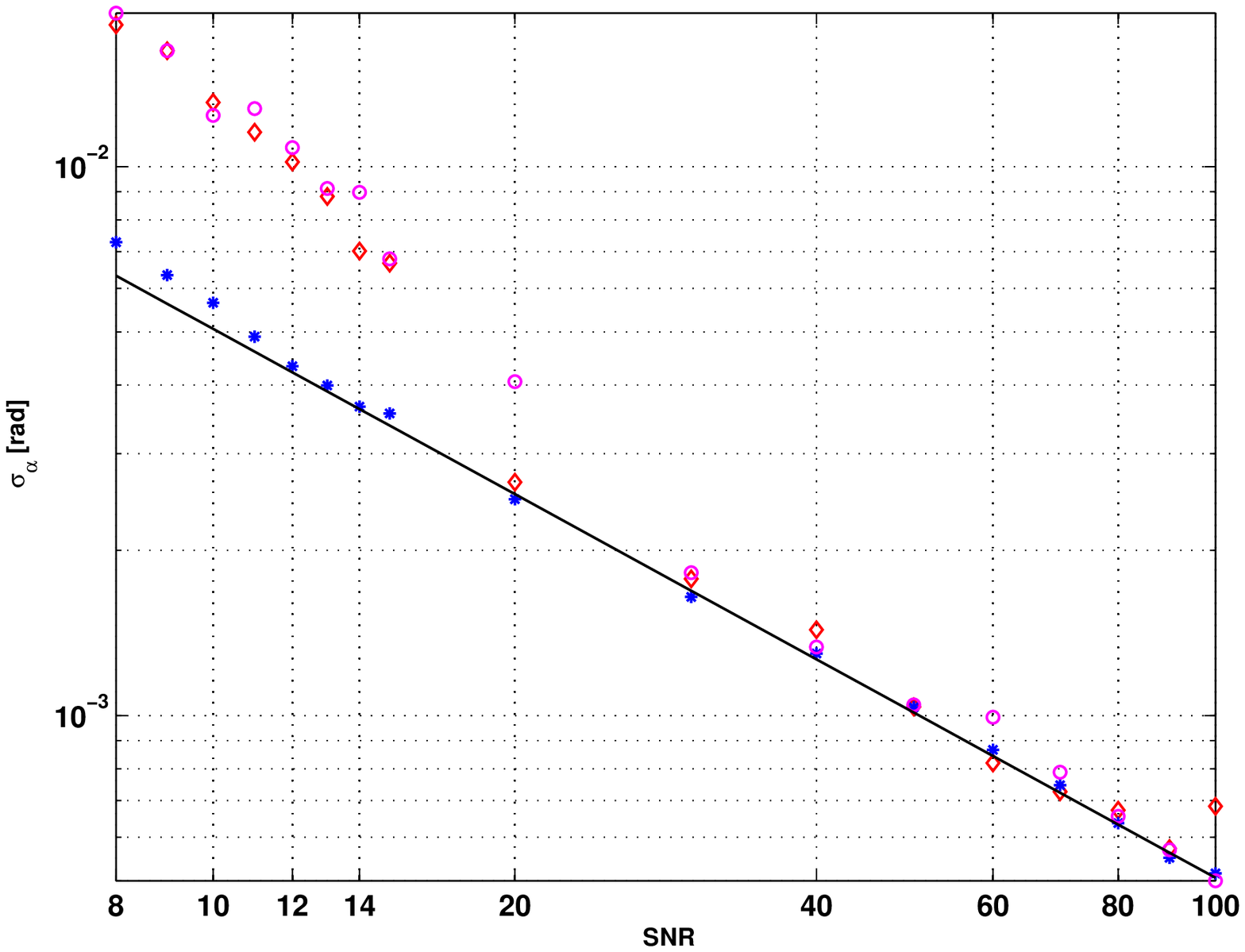} &
\includegraphics[scale=0.40]{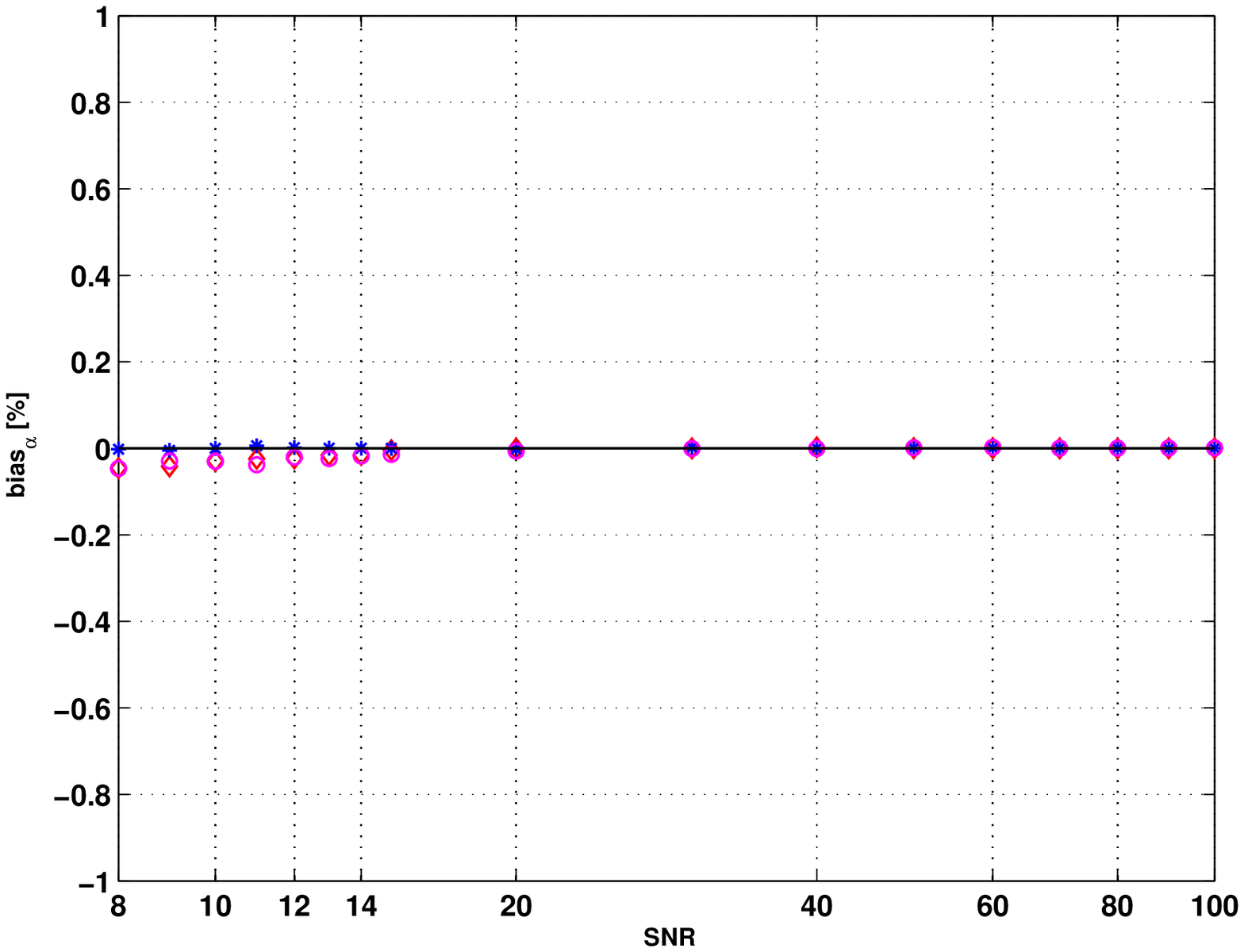}
\end{tabular}
\caption{\label{fig:sig2}
Standard deviations (left panels) and biases (right panels) of the ML estimators
of the intrinsic parameters as functions of the SNR
computed for the following three cases:
(i) interbinning and spline interpolation,
with the coarse grid of $\mathrm{MM}=\sqrt{3}/2$ (diamonds);
(ii) interbinning with the nearest neighbor interpolation,
with the fine grid of $\mathrm{MM}=\sqrt[3]{0.9}$ (stars);
(iii) interbinning with the nearest neighbor interpolation,
with the coarse grid of $\mathrm{MM}=\sqrt{3}/2$ (circles).
The continuous lines in the left panels are the Cram\`er-Rao bounds.}
\end{figure*}

In the third simulation we have studied another three specific cases:
(i) interbinning and spline interpolation,
with a coarse grid of $\mathrm{MM}=\sqrt{3}/2$;
(ii) interbinning with the nearest neighbor interpolation,
with a fine grid of $\mathrm{MM}=\sqrt[3]{0.9}$;
(iii) interbinning with the nearest neighbor interpolation,
with a coarse grid of $\mathrm{MM}=\sqrt{3}/2$.
The results of the simulation are presented in Fig.\ \ref{fig:sig2}.
This simulation shows that with a sufficiently fine grid
even the use of the least accurate (but the fastest) resampling
by the nearest neighbor interpolation
leads to the rms errors of the intrinsic parameters
very close to the Cram\'er-Rao bounds.

\begin{acknowledgments}

The contributions of K.\ M.\ Borkowski, P.\ Jaranowski, A.\ Kr\'olak,
and M.\ Pietka were supported in part by the MNiSzW grants
nos.\ 1 P03B 029 27 and N N203 387237.
A.K.\ would like to acknowledge hospitality
of the Max Planck Institute for Gravitational Physics in Hannover, Germany,
where part of this work was done. We would also like to thank
Holger Pletsch for discussions and helpful remarks.

\end{acknowledgments}

\end{document}